\documentclass[iop,apj,tighten]{emulateapj}
\usepackage{apjfonts}
\usepackage{graphicx}
\usepackage{amsmath,amstext}
\usepackage{tikz}
\usepackage{enumitem}
\usepackage[flushleft]{threeparttable}
\usepackage{color}
\usepackage{hyperref}
\usepackage{natbib}
\usepackage[titletoc,title]{appendix}
\definecolor{darkblue}{rgb}{0.0,0.0,0.4}
\hypersetup{colorlinks,
            linkcolor=darkblue,urlcolor=darkblue,
            citecolor=darkblue}

\newcommand{\angstrom}{\textup{\AA}\,}
\newcommand{\lya}{Ly$\,\alpha$ }

\shorttitle{Giant Ly-alpha Nebulae around Quasars}
\shortauthors{Borisova et al.}

\begin{document}

\title{Ubiquitous Giant \lya Nebulae around the brightest Quasars at $z\sim3.5$ revealed with MUSE \footnotemark[*]}

\author{Elena~Borisova\altaffilmark{1}, Sebastiano~Cantalupo\altaffilmark{1}, Simon~J.~Lilly\altaffilmark{1}, Raffaella~A.~Marino\altaffilmark{1}, Sofia~G.~Gallego\altaffilmark{1}, 
Roland~Bacon\altaffilmark{2}, 
Jeremy~Blaizot\altaffilmark{2}, 
Nicolas~Bouch\'e\altaffilmark{3}, 
Jarle~Brinchmann\altaffilmark{4,5},
C.~Marcella~Carollo\altaffilmark{1},
Joseph~Caruana\altaffilmark{7,8}, 
Hayley~Finley\altaffilmark{3},
Edmund~C.~Herenz\altaffilmark{6}, 
Johan~Richard\altaffilmark{2}, 
Joop~Schaye\altaffilmark{4}, 
Lorrie~A.~Straka\altaffilmark{4}, 
Monica~L.~Turner\altaffilmark{9}, 
Tanya~Urrutia\altaffilmark{6}, 
Anne~Verhamme \altaffilmark{2},
Lutz~Wisotzki\altaffilmark{6}
}

\affiliation{$^1$Institute for Astronomy, ETH Zurich, Wolfgang-Pauli-Strasse 27, 8093 Zurich, Switzerland}
\affiliation{$^2$Univ Lyon, Univ Lyon1, Ens de Lyon, CNRS, Centre de Recherche Astrophysique de Lyon UMR5574, F-69230, Saint-Genis-Laval, France}
\affiliation{$^3$Institut de Recherche en Astrophysique et Plan\'etologie (IRAP), CNRS, 9 avenue Colonel Roche, 31400 Toulouse, France}
\affiliation{$^4$Leiden Observatory, Leiden University, P.O. Box 9513, 2300 RA Leiden, the Netherlands}
\affiliation{$^5$Instituto de Astrof\'isica e Ci\^encias do Espa\c co, Universidade do Porto, CAUP, Rua das Estrelas, 4150-762 Porto, Portugal}
\affiliation{$^6$Leibniz-Institut f\"ur Astrophysik Potsdam (AIP), An der Sternwarte 16, 14482 Potsdam, Germany}
\affiliation{$^7$Department of Physics, University of Malta, Msida MSD 2080, Malta}
\affiliation{$^8$Institute of Space Sciences \& Astronomy, University of Malta, Msida MSD 2080, Malta}
\affiliation{$^9$MIT-Kavli Center for Astrophysics and Space Research, Massachusetts Institute of Technology, 77 Massachusetts Ave., Cambridge, MA 02139, USA}

\email{elena.borisova@phys.ethz.ch}

\begin{abstract}

Direct Lyman$\,\alpha$ imaging of intergalactic gas at $z\sim2$ has recently revealed giant cosmological structures around quasars,
e.g. the Slug Nebula \citep{C2014Natur}. Despite their high luminosity, the detection rate of such systems in narrow-band and spectroscopic surveys is less than 10\%, possibly encoding crucial information on the distribution of gas around quasars and the quasar emission properties. In this study, we use the MUSE integral-field instrument to perform a blind survey for giant \lya nebulae around 17 bright radio-quiet quasars at $3<z<4$ that does not suffer from most of the limitations of previous surveys. After data reduction and analysis performed with specifically developed tools, we found that \emph{each} quasar is surrounded by giant \lya nebulae with projected sizes larger than 100 physical kpc and, in some cases, extending up to 320 kpc. The circularly averaged surface brightness profiles of the nebulae appear very similar to each other despite their different morphologies and are consistent with power laws with slopes~$\approx-1.8$. The similarity between the properties of all these nebulae and the Slug Nebula suggests a similar origin for all systems and that a large fraction of gas around bright quasars could be in a relatively ``cold'' ($T\sim10^4$K) and dense phase. In addition, our results imply that such gas is ubiquitous within at least 50 kpc from bright quasars at $3<z<4$ independently of the quasar emission opening angle, or extending up to 200 kpc for quasar isotropic emission.

\end{abstract}


\maketitle
\newpage

\begin{deluxetable*}{rlcrccccclcl}{Hb}
\centering
\tablecolumns{8}
\tablewidth{0pt}
\tabletypesize{\footnotesize}
\tablecaption{Quasar sample and observation log}
\tablehead{
           \colhead{Number}                                & 
           \colhead{Quasar}                                & 
           \colhead{RA}                                    & 
           \colhead{DEC}                                   & 
           \colhead{$z_{cat}$\,\tablenotemark{a}}          &
           \colhead{$z_{sys}$\,\tablenotemark{b}}          &
           \colhead{$\lambda L_{1700}$\,\tablenotemark{c}} &
           \colhead{$i_{\mbox{AB}}$\,\tablenotemark{d}}    &
           \colhead{M$_i(z=2)$\,\tablenotemark{e}}         &
           \colhead{Class\,\tablenotemark{f}}              &
           \colhead{Seeing\,\tablenotemark{g}}             & 
           \colhead{Sky\,\tablenotemark{h}}                                   \\ 
           \colhead{}                                      & 
           \colhead{}                                      & 
           \colhead{(J2000)}                               & 
           \colhead{(J2000)}                               & 
           \colhead{}                                      &                           
           \colhead{}                                      &                           
           \colhead{[erg\,s$^{-1}$]}                       &                           
           \colhead{[$mag$]}                               &                           
           \colhead{[$mag$]}                               &                           
           \colhead{}                                      &                           
           \colhead{[arcsec]}                              & 
           \colhead{Conditions}                                     
}   
\startdata
 1 & CTS G18.01  &  00:41:31.4 & -49:36:11.9 & 3.240 & 3.207 & 6.50$\times 10^{46}$ & 16.621 & -30.22  & RQ & 1.08 & CL       \\
 2 & Q0041-2638  &  00:43:42.7 & -26:22:10.9 & 3.053 & 3.036 & 1.34$\times 10^{46}$ & 18.328 & -28.42  & RQ & 1.14 & WI-CL/TN \\
 3 & Q0042-2627  &  00:44:33.5 & -26:11:25.9 & 3.289 & 3.280 & 1.03$\times 10^{46}$ & 18.663 & -28.23  & RQ & 1.18 & WI-PH    \\
 4 & Q0055-269   &  00:57:58.1 & -26:43:15.8 & 3.662 & 3.634 & 3.33$\times 10^{46}$ & 17.475 & -29.52  & RQ & 1.02 & PH/CL    \\
 5 & UM669       &  01:05:16.7 & -18:46:41.9 & 3.037 & 3.021 & 2.06$\times 10^{46}$ & 17.811 & -28.92  & RQ & 1.31 & CL/PH    \\
 6 & J0124+0044  &  01:24:04.0 &  00:44:33.5 & 3.810 & 3.783 & 1.88$\times 10^{46}$ & 18.099 & -28.98  & RQ & 0.82 & PH       \\
 7 & UM678       &  02:51:40.4 & -22:00:28.3 & 3.205 & 3.188 & 1.72$\times 10^{46}$ & 18.014 & -28.81  & RQ & 0.72 & PH       \\
 8 & CTS B27.07  &  04:45:33.1 & -40:48:42.8 & 3.270 & 3.132 & 1.90$\times 10^{46}$ & 17.978 & -28.82  & RQ & 0.59 & PH       \\
 9 & CTS A31.05  &  05:17:42.1 & -37:54:45.9 & 3.020 & 3.020 & 1.98$\times 10^{46}$ & 17.824 & -28.91  & RQ & 0.72 & CL       \\
10 & CT 656      &  06:00:08.7 & -50:40:30.1 & 3.130 & 3.125 & 2.83$\times 10^{46}$ & 17.549 & -29.24  & RQ & 0.70 & CL       \\
11 & AWL 11      &  06:43:26.9 & -50:41:12.9 & 3.090 & 3.079 & 1.62$\times 10^{46}$ & 18.078 & -28.69  & RQ & 0.63 & PH       \\
12 & HE0940-1050 &  09:42:53.6 & -11:04:26.0 & 3.093 & 3.050 & 6.19$\times 10^{46}$ & 16.630 & -30.12  & RQ & 0.74 & CL       \\
13 & BRI1108-07  &  11:11:13.7 & -08:04:03.0 & 3.910 & 3.907 & 1.74$\times 10^{46}$ & 18.312 & -28.86  & RQ & 0.98 & TK/TN    \\
14 & CTS R07.04  &  11:13:50.1 & -15:33:40.2 & 3.370 & 3.351 & 2.78$\times 10^{46}$ & 17.601 & -29.36  & RQ & 0.94 & TN       \\
15 & Q1317-0507  &  13:20:29.8 & -05:23:34.2 & 3.700 & 3.701 & 3.20$\times 10^{46}$ & 17.525 & -29.51  & RQ & 0.94 & CL       \\
16 & Q1621-0042  &  16:21:16.7 & -00:42:48.2 & 3.700 & 3.689 & 4.31$\times 10^{46}$ & 17.079 & -29.95  & RQ & 0.85 & TK       \\
17 & CTS A11.09  &  22:53:10.7 & -36:58:15.9 & 3.200 & 3.121 & 2.07$\times 10^{46}$ & 17.815 & -28.98  & RQ & 0.76 & CL       \\
R1 & PKS1937-101 & 19:39:57.4  & -10:02:39.9 & 3.787 & 3.769 & 7.27$\times 10^{46}$ & 16.727 & -30.35  & RL & 0.75 & CL       \\
R2 & QB2000-330  & 20:03:24.1  & -32:51:45.9 & 3.783 & 3.759 & 4.26$\times 10^{46}$ & 17.302 & -29.77  & RL & 0.96 & CL

\enddata
\tablenotetext{a}{Taken from the catalog~\citet{Veron-Cetty}.} 

\tablenotetext{b}{Measured from MUSE spectra from the peak of the quasar C\,IV emission and correcting for luminosity-dependent velocity shifts using \citet{Schen2016_civ}. The $1\sigma$ level of the intrinsic uncertainty of the C\,IV correction relative to the systemic redshift is $\sim415$km\,s$^{-1}$ ($\Delta Z\sim0.006-0.007$).}

\tablenotetext{c}{Specific monochromatic continuum luminosity in the observed frame used to compute the correction for the estimated C\,IV redshift in \citet{Schen2016_civ}.}
                  
\tablenotetext{d}{Computed from MUSE datacubes with circular aperture photometry using a radius of 3 arcsec and assuming a SDSS i-band filter. No correction for galactic absorption has been applied.}

\tablenotetext{e}{Absolute i-band magnitude normalized at $z=2$ using \citet{Ross+2013}.}

\tablenotetext{f}{RQ - radio-quiet quasar; RL - radio-loud quasar. Classification is based on radio fluxes measurements from \citet{2001ApJ...555..625C} and \citet{1998AJ....115.1693C} for RQ and RL respectively.}

\tablenotetext{g}{Seeing measured in the combined 1 hour datacubes as the FWHM of a Gaussian profile.}

\tablenotetext{h}{Sky conditions during the night of observations. The meaning of the labels is the following: PH-photometric night; CL-Clear night; WI-Strong winds; TN-thin clouds; TK-thick clouds.}

\label{tab1}
\end{deluxetable*}

\section{Introduction}
\label{intro}

\renewcommand*{\thefootnote}{\fnsymbol{footnote}}
\footnotetext[1]{Based on observations made with ESO Telescopes at the Paranal Observatory under programs 094.A-0396, 095.A-0708, 096.A-0345, 094.A-0131, 095.A-0200, 096.A-0222}
\renewcommand*{\thefootnote}{\arabic{footnote}}
\setcounter{footnote}{0}

The Intergalactic Medium (IGM) plays a central role in our understanding of how structures form and evolve in the Universe. Our standard cosmological model predicts that the bulk of the baryons in the Universe should reside in  a "Cosmic Web"  of intergalactic filaments ~\citep{Bond_etal_CWeb,Fukugita1998,Dave_baryons} that directly trace the underlying dark matter distribution but are too diffuse to form stars. These filaments represent a rich reservoir of pristine gas that drives galaxy formation and evolution, especially in the early Universe (e.g., \citealp{Keres2005, Dekel+2009,Voort2011,F+2011}).

The diffuse nature of the IGM represents a challenge for observational studies. One of the most efficient ways to trace the distribution of intergalactic gas and to study its physical conditions is through hydrogen \lya absorption line studies using the spectra of distant quasars. Unfortunately, the sparseness of these one-dimensional probes typically precludes direct constraints on the three-dimensional morphology and small scale properties of individual intergalactic filaments. As a consequence, the possible role of gas filaments in feeding galaxies and quasars in the early Universe is still poorly constrained.

Direct imaging of at least the densest parts of the IGM has in recent years become a concrete possibility thanks to improved instrumentation and new observational probes such as 
quasar fluorescent \lya emission. Following the early prediction of \citet{Hogan_Weyman,Gould,HaimanRees} and more detailed radiative transfer studies focusing also on quasar illumination 
\citep{C2005,Kollmeier2010}, fluorescent \lya surveys were successfully carried out using custom-built narrow-band (NB) filters on 8-meter class telescopes and ultra-deep integration on hyper-luminous and radio-quiet quasars at $z\sim2$ \citep{C2012, Hennawi2015,C2014Natur}. These surveys have revealed dense and compact intergalactic clouds with very little star formation ("dark galaxies"), circum-galactic streams around star forming galaxies \citep{C2012} and two giant nebulae with sizes of about 460\,physical kpc (pkpc) \citep{C2014Natur} and 350\,pkpc \citep{Hennawi2015} in proximity of the quasars. The typical detection rate of giant nebulae (i.e., with projected sizes larger than 100\,pkpc) around radio-quiet quasars in these NB surveys is less than 10\% considering both deep LRIS/Keck data \citetext{Cantalupo et al., in prep.} and shallower surveys using GMOS \citep{Fab2016}. This low detection rate is also found by other independent surveys using broader NB filters on LRIS/Keck \citep{Martin2014}.

To date the two giant nebulae detected by \citet{C2014Natur} (the "Slug" Nebula) and \citet{Hennawi2015} (the "Jackpot" Nebula) and the extended emission reported by \citet{Martin2014} 
are the only radio-quiet \lya nebulae with sizes significantly larger than 100\,pkpc. Other NB surveys \citep[e.g.][]{Hu1991,Fab2016} and spectroscopic observations (e.g., \citealp{ Christensen2006,North2012,HennawiX2013, Herenz2015}) have either not detected any extended emission at all or detected emission on much smaller scales ($\lesssim$ 50-60\,pkpc) was found in about 50\% 
of the cases\footnote{The only exception is the shallow NB observations of \citealp{Bergeron1999}, which detected emission extending about 100\,pkpc around the J2233-606 radio-quiet quasar in the parallel HDF-S field. However preliminary but deeper GMOS NB imaging does not currently confirm such extended emission \citep{Fab2016}. 
In addition, the local radio-quiet quasar at z$\sim0.064$ MR\,2251-178 also shows extended emission in H$\alpha$ and [OIII] on scales larger than 100 kpc \citep[e.g.][]{Bergeron1983,Shopbell1999,Kreimeyer2013}. However, because of the very different redshift and the lack of \lya emission information for this nebula we have not included this object in our current comparison sample.
}.
In contrast, the detection rate of \lya nebulae with sizes of about 100\,pkpc is larger than 80\% for NB imaging and spectroscopy around radio-loud quasars \citep[e.g][]{Heckman1991a,Roche-RL}, and the most luminous and distant radio galaxies are almost always associated with large \lya nebulae with sizes of up to 200\,pkpc \citep[e.g.][]{McCarthy1993, Reuland2003}. However, the much broader \lya line profiles of the nebulae associated with these radio-loud sources (with a line full width half maximum FWHM$\,>\,$1000\,km\,s$-1$), the alignment between the extended \lya emission and the radio-loud lobes, and the higher metallicities all suggest a different origin with respect to radio-quiet systems, e.g. outflows rather than intergalactic filaments, at least for the inner parts of the \lya emission (\citealp{Heckman1991b,VM2003,Humphrey2007}, but see also \citealp{Villar-Martin}).

In principle, the detection rate of giant fluorescent nebulae around quasars should depend on both the presence of intergalactic cold (T$\sim10^4$ K) gas around the quasars and on the ``illumination'' provided by these bright UV sources. Absorption line studies using quasar pairs have found a high covering fraction ($\sim60$\%) of optically thick gas at projected distances of 200\,pkpc from the quasars \citep[e.g.][]{Prochaska2013,Heyley2014}, suggesting the presence of large amounts of cold gas around these quasars. Combining the constraints from absorption and emission studies\footnote{Note, however that quasar pairs, selected for absorption studies, are typically less luminous than the individual quasars that have been studied in emission.}, one might be  tempted to interpret  the low detection rate of giant emitting nebulae as a consequence of a small opening angle of the quasar radiation "beam" together with an anisotropic distribution of the "cold" gas.
 
In reality, however, many factors related to the observational techniques may play a role in determining the detection rate of giant nebulae. For instance, NB imaging relies on the availability of accurate systemic redshifts for the quasars and in some cases the nebular \lya line may fall at the edge or outside of the filter (filter losses). Although in some cases accurate redshifts from near-infrared spectroscopy are available (e.g. for the Slug Nebula, \citealp{C2014Natur}), the majority of quasar redshifts (e.g. from SDSS) are typically estimated from broad emission lines such Mg\,II that have an error in velocity comparable to the central part of the NB filters themselves \citep{2010MNRAS.405.2302H}. Filter losses could be particularly relevant for kinematically narrow nebulae while broader nebulae, such as radio-loud systems, would be less affected. Long-slit spectroscopic surveys, on the other hand, can only cover a small part of the area around the quasars (i.e., they suffer from slit losses) and all giant nebulae discovered so far are clearly asymmetric \citep{C2014Natur, Martin2014, Hennawi2015}. Moreover, in order to discover faint and extended nebulosities, it is necessary to properly remove the point spread function (PSF) associated with the hyper-luminous quasars, a task that is particularly difficult for some instruments (e.g. LRIS/Keck) and in general for NB imaging (PSF losses). Finally, sensitivity is certainly a factor but likely less relevant because the giant nebulae discovered so far have been extremely bright (total \lya luminosities $\approx10^{44}-10^{45}\,$erg$\,$s$^{-1}$). 
 
In this study, we exploit the power of the new Multi Unit Spectroscopic Explorer \citep[MUSE;][]{MUSEref} which is an integral field spectrograph on the Very Large Telescope (VLT) to overcome these technical limitations of previous NB and spectroscopic surveys. MUSE is the ideal instrument to search for giant \lya nebulae around quasars thanks to its large field of view (1'$\times$1') and because by design it does not suffer from either filter losses or slit losses. Also, the large number of spatial and spectral elements allows for a very accurate quasar PSF estimation and removal. Because accurate systemic redshifts are not needed for spectroscopic surveys, any quasar with \lya redshifted between the blue and red edges of the MUSE wavelength range ($2.9<z<6.5$) can be observed. In this first exploratory study as a part of the MUSE Guaranteed Time Observations (GTO), we selected 12 of the brightest  radio-quiet quasars in the Universe at $z\approx3.2$ (to maximize throughput) and complemented these with 7 other quasars, of which 5 are radio-quiet, at $z\approx3.7$ that have been observed with MUSE as part of a different GTO program. As we will show in the following sections, the picture emerging from these MUSE observations is very different than that based on previous surveys,
in that giant nebulae with sizes larger than 100\,pkpc are found around essentially every radio-quiet quasar.  

We present our results in the following order. In Section~\ref{datasec} we describe our target selection, observational strategy and basic data reduction steps. We give a detailed overview of the taken steps to search for extended \lya emission in Section~\ref{detection}. In Section~\ref{results} we describe the observational properties of the detected giant \lya nebulae. In Section ~\ref{discussion}, we compare our results with previous studies and discuss the implications of our findings. The summary and conclusions are given in Section~\ref{conclude}.

Throughout the paper, we assume a $\Lambda$CDM cosmology with $\Omega_{m}=0.3,\Omega_{\Lambda}=0.7$, and $h=70$~km s$^{-1}$. The units of size are pkpc. One arcsec at $z\approx3.1$ and $z\approx3.7$ correspond to $7.6\,$pkpc and $7.2\,$pkpc respectively.

\section{Observations and Data Reduction}
\label{datasec}

\subsection{Sample selection}

Our total sample of 17 radio-quiet quasars, complemented by 2 radio-loud systems (see Table~\ref{tab1}), is composed of two subsamples reflecting two different 
MUSE GTO programs (094.A-0396, 095.A-0708, 096.A-0345 PI:\,S.\,Lilly; 094.A-0131, 095.A-0200, 096.A-0222 PI:\,J.\,Schaye). The main subsample of 12 radio-quiet quasars has been constructed specifically for this study (094.A-0396, 095.A-0708, 096.A-0345) using the catalog of \citet{Veron-Cetty} and selecting the brightest radio-quiet quasars known in the redshift range $3.0<z<3.3$ to maximize MUSE throughput and minimize redshift-dimming of the surface brightness (SB). We have removed quasars in proximity of bright stars and fields with high galactic extinction. Among the remaining quasars, we have then selected objects with available UVES spectroscopy to maximize "transverse" science projects within the MUSE GTO program. 

The second subsample of 5 radio-quiet and 2 radio-loud quasars is part of a MUSE GTO program to study the connection between absorption line systems in quasar spectra and 
emission-line galaxies (094.A-0131, 095.A-0200, 096.A-0222). The selection criteria of this program required quasars at higher redshift, i.e.\ $z\approx3.6-4.0$ in order to maximize the available path-length of the \lya forest within the MUSE wavelength range. Moreover, these quasars have very deep UVES spectroscopy. Apart from the different redshift, the 5 radio-quiet quasars of this subsample are very similar to the lower redshift objects in terms of luminosity. Although not originally part of this study, we have also included 2 radio-loud quasars from this subsample out of the 3 observed so far. In particular, we have excluded one radio-loud quasar (B1422+2309) from our sample because it is gravitationally lensed \citep{Lens1992} into multiple components making PSF subtraction challenging and because the field is very crowded with foreground galaxies.

\subsection{Observational strategy}

The three giant nebulae discovered hitherto with NB imaging \citep{C2014Natur,Martin2014,Hennawi2015} are all characterized by bright extended emission with surface brightness values larger than 10$^{-17}$~erg\,s$^{-1}$\,cm$^{-2}$\,arcsec$^{-2}$, a value that is easily reachable within 1 hour of integration time with MUSE. Therefore, for this exploratory survey (094.A-0396, 095.A-0708, 096.A-0345), we use a total exposure time of 1 hour for each quasar split into 4x900\,s exposures. Between each individual exposure we rotate the field of view (FoV) by 90 degrees and apply a small random dithering pattern of less than 1 arcsec. Before starting each observation, we offset the quasar position by a few arcsec away from the center of the FoV to avoid regions with higher systematics. A similar strategy has been applied to every quasar field observed as a part of the MUSE GTO. Some of the quasars observed as a part of programs 094.A-0131, 095.A-0200, and 096.A-0222 have a total integration time longer than 1 hour. In order to keep our sample homogeneous in depth, we have selected only the first 4x900\,s exposures for these quasars.

Data was collected with the MUSE/VLT instrument \citep{MUSEref} between 19 September 2014 and 9 November 2015. All but two observation blocks (OBs) were executed continuously during the same nights. The seeing varied in the range 0.59-1.31 arcsec (FWHM of the Gaussian at 7000\,\angstrom, measured in the combined 1 hour datacubes). The information about the quasar fields is summarized in Table~\ref{tab1}. Together with name and coordinates, we provide: 
i) the original redshift from the catalogue, ii) our systemic redshift estimate based on the C\,IV emission corrected for blueshift according to the quasar luminosity as in \citet{Schen2016_civ}, 
iii) the luminosity at 1700\,\angstrom rest-frame, iv) the i-magnitude and corresponding absolute magnitude normalized to $z=2$ as in \citet{Ross+2013}, and v) the radio class, seeing and sky conditions during each night.

\subsection{Data Reduction}

We used the standard MUSE pipeline v1.0 \citep[][Weilbacher in prep.]{MUSEpipe2012,MUSEpipe2014} for the basic steps of the data reduction with the default (recommended) parameters. For each of the individual exposures we performed bias subtraction, flat-fielding, twilight and illumination correction, and wavelength calibration. Sky subtraction was not performed with the pipeline but was done at a later stage with custom developed software as discussed below. The response curve and telluric correction were obtained from one of the spectrophotometric standards observed during the same night except for field  \#17 where we had to use a standard star from the previous night. Finally, flux calibrated data was drizzled onto a 3D grid using the information from geometry and astrometry tables in order to produce the final datacubes. 

It is known that MUSE cubes reduced with the standard pipeline might have small astrometric offsets in the coordinate system because of a small "derotator wobble" \citep{MUSE-HDFS}. To correct for this effect, 
we registered the datacubes using the position of point sources in MUSE white-light images (obtained by collapsing the datacubes along the wavelength direction) in different exposures.

The final steps of the data reduction were performed with custom tools for flat-fielding correction and sky-subtraction that are part of the CubExtractor package \citetext{Cantalupo, in prep.} 
and that have been specifically developed to improve data quality for the detection of faint and diffuse emission in MUSE datacubes. In particular, the flat-fielding correction is performed as a self-calibration on each individual datacube using the sky-continuum and the sky-lines as a spatially uniform source to re-calibrate each individual slice (part of an Integral Field Unit, IFU) and IFU as a function of wavelength. Sources are masked with an iterative procedure to ensure that self-calibration errors are minimized. With this procedure, the typical striped-pattern of broad-band and NB images of MUSE cubes is totally removed and any visible residual is typically at a level much less than 0.1\% of the sky (a more detailed description of this procedure, called CubeFix, will be presented in Cantalupo, in prep.). In some rare cases, there are not enough spatial and spectral elements in a slice to find a suitable correction or there is clear variation in a single slice that cannot be corrected with a simple rescaling factor. In these cases, we mask the volume pixels (voxels) in the datacube corresponding to the slices or region without a proper correction factor. Because of dithering and FoV rotation, these masked regions do not typically affect the final combined datacube. 

Sky-subtraction is then performed on each individual, flat-field corrected cube using CubeSharp \citetext{Cantalupo, in prep.}. CubeSharp uses a local and flux-conserving procedure to empirically correct the sky line spread function (LSF) and therefore remove sky lines minimizing the residuals due to LSF shifts and variation across the MUSE FoV - the major sources of systematic errors in MUSE cubes. Because the algorithm is flux-conserving by design, no residuals are introduced when sky-subtraction is performed with CubeSharp.

Finally, the corrected and sky-subtracted cubes are combined using an average $3\sigma$-clipping algorithm. After this first iteration, a white-light image is created and continuum sources are identified using CubExtractor \citetext{Cantalupo, in prep.} (see Section ~\ref{cubex}). Using the positions and spectra of continuum sources from the combined cube, another iteration of CubeFix and CubeSharp is performed on individual exposures to improve the removal of self-calibration effects. Typically one iteration is sufficient to substantially improve the data reduction process before the individual cubes are combined again.

\begin{figure*}[ht]
\centering
\includegraphics[width=0.88\textwidth]{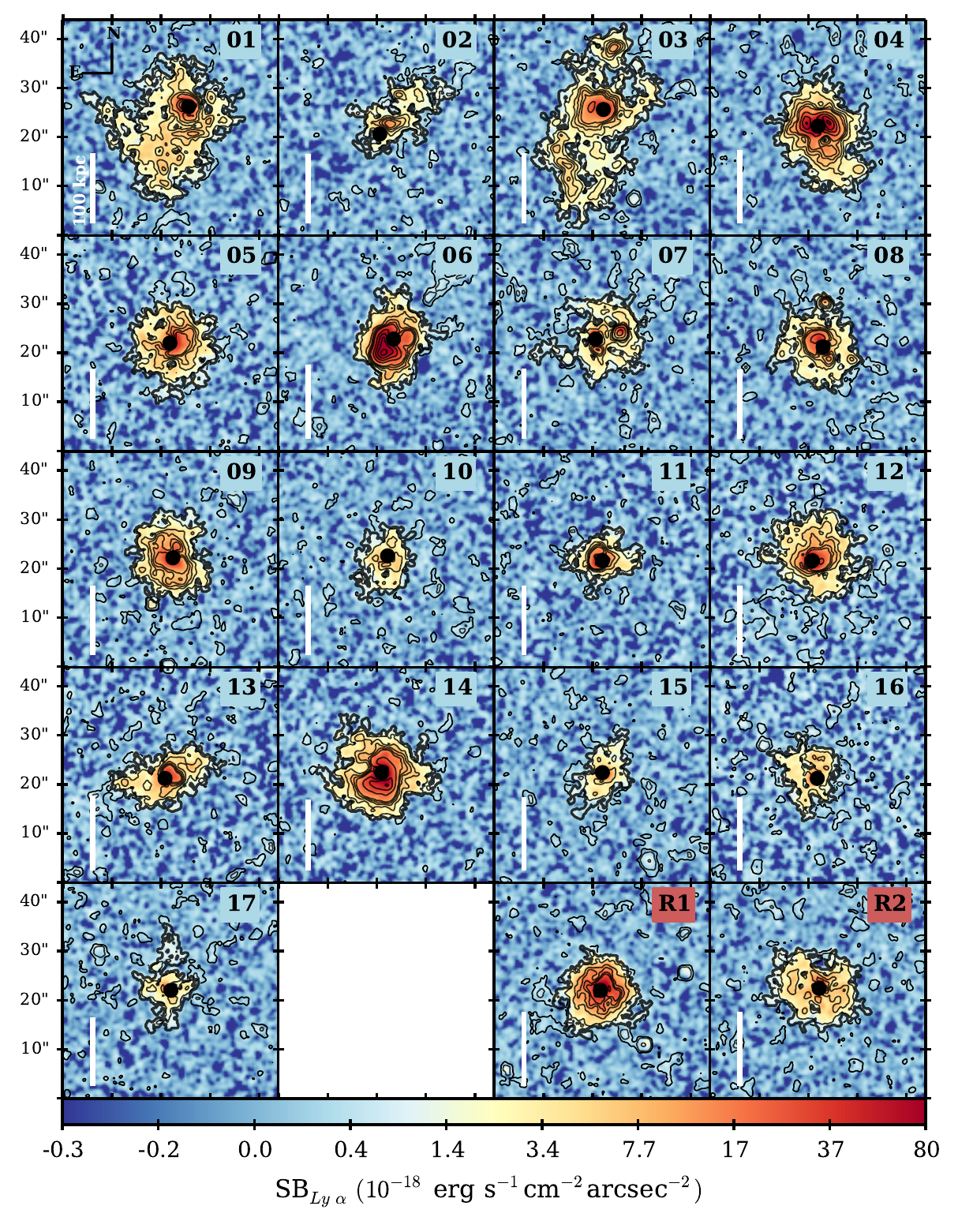}
 \caption{"Optimally-extracted" \lya images from PSF and continuum subtracted MUSE datacubes obtained with CubExtractor for each quasar observed in this study. Each image has a linear projected size of 44 arcsec, and the original position of the quasar is marked by a black dot. The white bar indicates a physical scale of 100\,kpc. 
 The images have been produced by collapsing the datacube voxels associated with the CubExtractor three-dimensional segmentation maps (the "3D-mask") along the wavelength direction (see ~\ref{results-sub1}). The 3D-masks have been obtained with a signal-to-noise ratio (SNR) threshold of 2 per smoothed voxel as discussed in Section ~\ref{detection}. For display purposes, we have added - by means of the union operator - to the object 3D-mask one wavelength layer of the cube corresponding to the central wavelength of the nebulae. The spatial projection of the 3D-mask is indicated by the thick contours that typically correspond to a SB of about 10$^{-18}$ erg s$^{-1}$ cm$^{-2}$ arcsec$^{-2}$. The thin contours indicate the propagated SNR in the images. The two highest contour levels represent SNR=2 and SNR=4, while the other contours are separated by $\Delta$SNR=6. As is clear from this image, each field shows the presence of extended Ly$\alpha$ emission at a high significance level.}
 \label{fig1-nebulae}
\end{figure*}

\begin{figure*}[ht]
\centering
\includegraphics[width=0.92\textwidth]{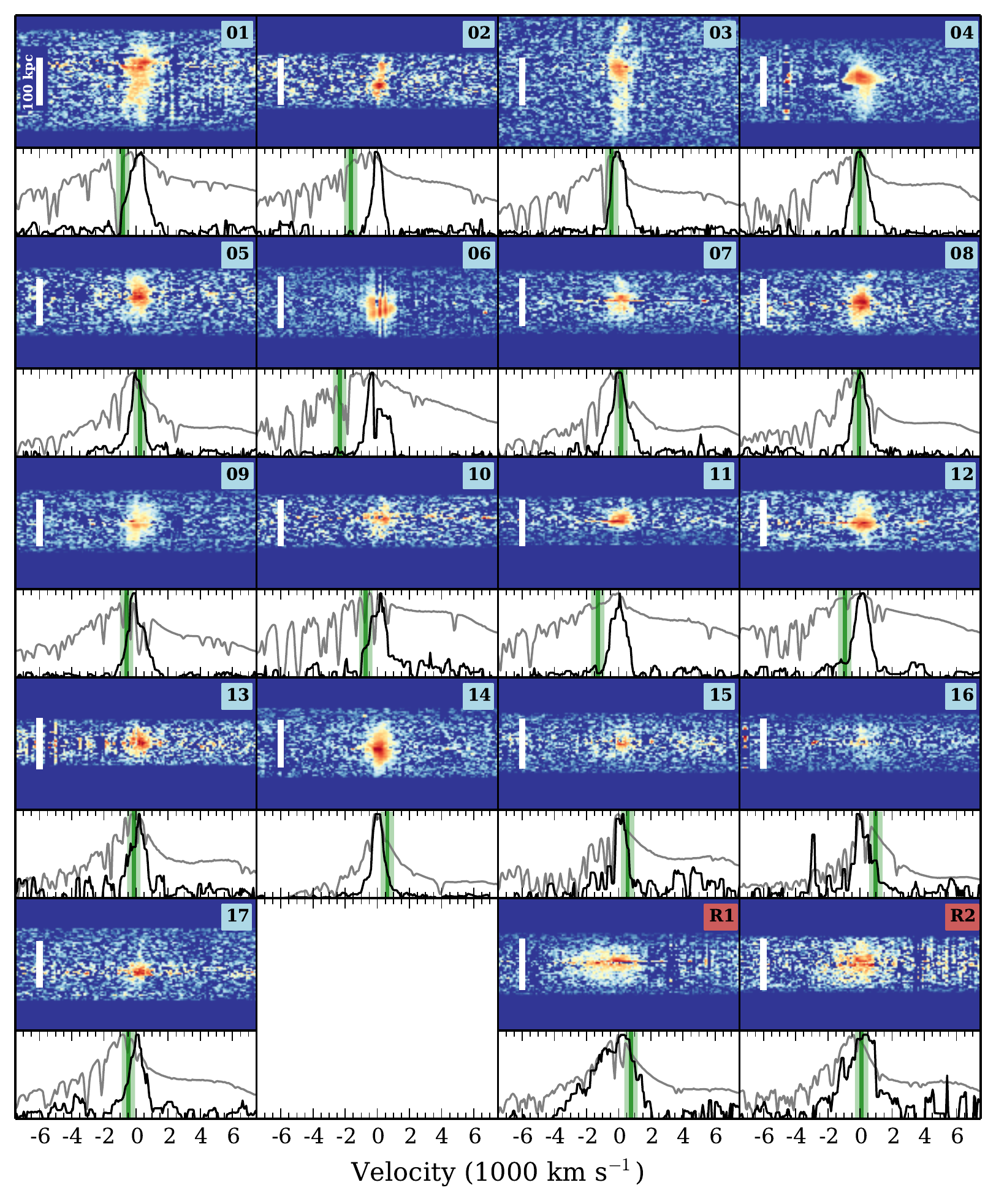}
 \caption{"Optimally-extracted" two-dimensional (upper panels) and one-dimensional (lower panels) \lya spectra obtained from MUSE datacubes with CubExtractor (Cantalupo, in prep.) for each
 observed quasar in this study. The two-dimensional spectra have been extracted using the pseudo-aperture defined by the spatial projection of the "3D-mask" (shown in Fig.~\ref{fig1-nebulae} as a thick contour)
 and integrating along the spatial x-axis direction. The vertical stripes in the nebula \#6 are due to residuals of the sky subtraction. The white bar shows a spatial scale of 100\,pkpc. The lower panels show the one-dimensional spectra of the nebulae (thick black lines)  obtained by integrating the two-dimensional spectra along the remaining spatial direction. For comparison, we overlay as a grey line the one-dimensional spectrum of the quasar integrated in a circular aperture with a radius of 3\,arcsec. Clearly, all detected radio-quiet nebulae show a \lya spectral shape very different from that of the quasar, confirming that the detected emission is not an artefact of the quasar PSF subtraction. The green lines indicate the estimated quasar systemic redshift using the blueshift-corrected C\,IV emission taking into account the quasar luminosity as in \citet{Schen2016_civ}. The shaded green area represents the 1$\sigma$ error associated with the systemic redshift calibration (415\,km\,s$^{-1}$).}
 \label{fig1-spec}
\end{figure*}

\section{Detecting extended \lya emission}
\label{detection}

In order to search for extended low surface brightness \lya emission around the quasars in our sample, we developed a common scheme which we describe below step by step. It was applied to each of the final combined cubes.

\subsection{Subtraction of the quasar PSF}

To reveal the presence of extended \lya in the vicinity of a quasar, we need to remove the contribution from the quasar PSF. The advantage of Integral Field Spectroscopy (IFS) is that we have images of the quasar and its surroundings at different wavelengths, including regions in the spectra for which we are sure that no extended line emission should be present.

We have explored two different approaches of quasar PSF estimation and removal. One way is to fit a known function, for example a Gaussian or Moffat profile. Unfortunately, the MUSE PSF in reconstructed and combined cubes is complex due to the nature of the instrument and the wings of the PSF profiles cannot be easily modelled by these simple analytic profiles. Indeed, both Gaussian and
Moffat fitting produced clear residuals of the quasar PSF in subtracted images at level that would preclude the detection of faint and extended \lya emission unrelated to the quasar PSF (see also \citealp{Christensen2006} and references therein).
 
The other method is purely empirical and uses the data itself to construct PSF images that are rescaled and subtracted at each wavelength layer\footnote{The algorithm, called CubePSFSub, is also part of the CubExtractor package (Cantalupo, in prep.)} (see also \citealp{Husemann2013} and \citealp{Herenz2015} for an other iterative empirical approach). In particular, for each wavelength layer we produce a pseudo-NB image with a spectral width of 150 spectral pixels ($\approx187\,\angstrom$) at the position of the quasar. This spectral width is a compromise between minimizing the PSF wavelength variations and maximizing the SNR in the empirical PSF images. 
The flux in each empirical PSF image is then rescaled assuming that the quasar is dominating the flux in the majority of the central 5$\times$5 pixels, i.e.\ $1\times1$\,arcsec$^2$. 
In particular, because the flux in individual pixels may be affected by cosmic rays or other artefacts, we compute the rescaling factor between the flux in each layer and the empirical PSF images using an averaged-sigma-clip algorithm. Once the empirical PSF image has been rescaled we cut from it a central circular region with a radius of about 5 times the seeing (masking any pixel with negative flux) and subtract this circular cut-out from the corresponding wavelength layer in the datacube. In some cases, the nebulae are so bright that their central parts will be visible at a low level even in a $\approx187\,\angstrom$-wide band. In this case, we iterate the PSF removal procedure, masking the wavelength region associated with the nebulae to avoid over-subtraction. 

This empirical PSF subtraction produces excellent results, especially on large scales around the quasar. However, we note that this method cannot provide any meaningful information on the central 1 arcsec region used for the PSF rescaling and that it is not able to treat blending of the quasar PSF with other nearby continuum sources, if present. For the purposes of this study, we will see that the method is adequate since the detected nebulae extend far beyond the central 1 arcsec region around the quasars. Moreover, we mask and remove continuum sources around the quasar as described below, so their residuals do not affect the results presented here. 

\subsection{Continuum source subtraction}

After quasar PSF subtraction has been performed, we remove any other possible continuum sources for each spaxel in the cube using a fast median-filtering approach based on the following method:
i) we first resample the spectrum of each spaxel into spectral regions (bins) with a size of 150 spectral pixels ($\approx187\,\angstrom$) using a median filter, 
ii) the resampled spectrum is then smoothed with a median filter with radius equivalent to 2 bins to minimize the effect of line features in individual spectral regions, 
iii) we subtract from each voxel in the cube the estimated continuum from the corresponding spaxel and spectral region. 
This procedure allows a fast and efficient removal of continuum sources (a full median-filter approach for each spaxel and each spectral pixel would be more computationally expensive). In some cases, stars or background galaxies with emission lines in their spectra are not properly removed by this procedure due to the large window size and negative or positive residuals are visible in the cube. However, these residuals do not affect our results because we mask any bright star or galaxy in the cube before extraction, as described below.

\subsection{3D detection and extraction}
\label{cubex}

The final step in our data analysis procedure is the detection and extraction of extended line emission from the reduced and processed cubes. For this task we use newly developed three-dimensional automatic extraction software called CubExtractor \citetext{Cantalupo, in prep.} based on a three-dimensional extension of the connected-labelling-component algorithm with union finding of classical binary image analysis (see, e.g. \citealp{Shapiro01}). In particular, after datacubes and associated variances are being filtered (smoothed), voxels above a given user-defined signal-to-noise ratio (SNR) threshold are connected and objects are detected if they contain a certain number of connected voxels above a user-defined threshold. Three-dimensional segmentation masks produced by CubExtractor are then used for photometry and in order to obtain several lower-dimensional projections of the extracted objects, such as optimally-extracted images, two-dimensional spectra, and velocity maps, as we will show in the next sections. 

Because detection and extraction is based on a SNR threshold, the characterization of the noise in MUSE datacubes is a crucial aspect of the process. A propagated variance, estimated for each individual step of the reduction process, is provided by the MUSE pipeline and propagated during flat-fielding and continuum subtraction with CubeFix and CubeSharp. The variance is also propagated during exposure combination. The variance associated with MUSE resampled datacubes by the pipeline is an underestimate of the true variance because resampling introduces correlated noise that cannot be easily captured by current detection and extraction algorithms. In order to include this extra source of noise in an approximate fashion and to obtain the right ``normalisation'' for the variance, 
we proceed in the following way: i) we estimate from the cube itself the "spatial" variance of the pixel flux for each wavelength layer (this estimate is essentially a single value for each layer and therefore does not contain information on the spatial variation of the noise),
ii) we compute for each wavelength layer the spatial average of the propagated variance obtained by the pipeline, iii) we rescale the propagated variance by a constant factor for each wavelength layer in order to match the average "spatial" variance estimated from the cube itself. The value of the rescaling factor is typically around 1.4.

Using an initial guess for the redshift of the quasar (from the original catalogue or using the location of C\,IV\,/1550$\,\angstrom$ and He\,II\,/1640$\,\angstrom$ lines), we extract various subcubes from the processed datacube with wavelength ranges capturing the expected \lya line, C\,IV\,/1550$\,\angstrom$ and He\,II\,/1640$\,\angstrom$ lines (the sizes of the subcubes correspond to $\sim15,000\,$km$\,$s$^{-1}$). Because we are only interested in extended emission, 
we use a large number of minimum connected voxels in CubExtractor for detection ($10,000$). Moreover, we apply before detection a spatial gaussian filtering with the $\sigma$ value of  0.5 arcsec (without smoothing in wavelength) to bring out extended but narrow features, and we use a minimum SNR of 2 with respect to the rescaled variance cubes as described above. This value typically corresponds to a surface brightness of about 10$^{-18}$ erg s$^{-1}$ cm$^{-2}$ arcsec$^{-2}$ for a 1 arcsec$^{2}$ aperture in a single wavelength layer (i.e., 1.25$\,\angstrom$) for our cubes. 

With this set of parameters and masking regions associated with bright continuum-sources or sky residuals, we always find one single detection in the expected Ly$\alpha$ wavelength range for every quasar field. As we will show in the next section, these detections are giant Ly$\alpha$ nebulae extending up to several tens of arcsec around the quasars.

\section{Results}
\label{results}

\subsection{100\% detection rate of giant Ly$\alpha$ nebulae}
\label{results-sub1}

As discussed in Section~\ref{detection}, we have detected an extended source in each individual quasar field around the expected wavelength for \lya emission. Each of these sources is characterized by more than $10,000$ connected voxels (or individual spatial and spectral elements in MUSE cubes) with a SNR greater than 2 - after smoothing - and the confidence level of the detections, as we will show in this Section, is very high.

In Fig.~\ref{fig1-nebulae}, we present the "optimally extracted images" of the detected objects in each MUSE cube. Each image has a linear size of 44 arcsec and the original position of the quasar is marked by a black dot. 
These images have been obtained using the three-dimensional segmentation mask (called 3D-mask here for simplicity) associated with the detection object. We note that the 3D-mask defines a three-dimensional iso-SNR surface in the cube (after spatial smoothing as described in Section~\ref{detection}) and therefore is ideal to obtain images and spectra with maximal SNR after collapsing one of the spatial or the spectral dimensions. 
In particular, the images presented in Fig.~\ref{fig1-nebulae} have been obtained by selecting all voxels in the PSF-subtracted and continuum-subtracted MUSE cubes, using corresponding 3D-masks of each nebula, and integrating their fluxes along the wavelength direction. These images can also be thought of as pseudo-NB images where the width of the filter is adjusted for each
spaxel to maximize the SNR of the objects. The width of this pseudo-filters vary from one layer (typically at the edges of the object) to a few tens of layers in the brightest or kinematically broader parts of the sources. 
The projection of the 3D-mask on the plane of the sky is indicated by the thick contours overlaid on the images in Fig.~\ref{fig1-nebulae} and typically corresponds to a SB of about 10$^{-18}$ erg s$^{-1}$ cm$^{-2}$ arcsec$^{-2}$. We note that this deep sensitivity level - obtained in a single hour of MUSE observation - is comparable to the sensitivity of a 20h observation with a 40$\,\angstrom$ NB filter on an 8-meter telescope \citep[e.g.][]{C2012}.
For display purposes, we have added - by means of the union operator - to the 3D-mask one wavelength layer of the cube corresponding to the central wavelength of the nebulae. The voxels associated with this layer are shown outside of the thick contours in Fig.~\ref{fig1-nebulae}.

We stress that these "optimally extracted images" are quite different from standard NB (see Appendix~\ref{appA}) or broad-band images  - with which the reader may be more familiar - because the number of wavelength layers (and therefore the flux and corresponding noise) contributing to the image depends on spatial position and is thus correlated with the SNR. This gives us the unique possibility of capturing in a single image, at the best SNR threshold, the surface brightness values for both kinematically narrow and broad features that would have been either lost in the noise or underestimated in a NB image with a single width. As a consequence, the images in Fig.~\ref{fig1-nebulae} have a much larger dynamic range, despite the short integration time, with respect to a standard image. One possible drawback of this approach, however, is that the image noise cannot be simply estimated visually as the noise will be correlated with the number of layers and therefore with the flux and spatial position. To help the reader to visualize the true noise in these images, we have therefore estimated the noise and the SNR for each pixel \emph{in the image} by variance propagation (using the same rescaling factor derived in the CubExtractor run) taking into account the number of layers contributing to each pixel. 
The thin contours in Fig.~\ref{fig1-nebulae} show the \emph{image} SNR contours using these propagated variances and the integrated flux. The first contour corresponds to a SNR of 2, the second one to 4 and further steps between the contours to $\Delta$SNR=6. These contours are the closest representation to the true image noise as they are, ideally, independent of the number of layers contributing to each pixel (of course, this would not be true in case of significant correlated noise in the wavelength direction). 

In Fig.~\ref{fig1-spec}, we show different projections of the PSF and continuum subtracted MUSE datacubes using the CubExtractor 3D-masks, i.e.\ the "optimally extracted" two-dimensional and one-dimensional spectra. 
In each case, we have used as an aperture the spatial projection of the 3D-masks, i.e.\ the thick contours shown in Fig.~\ref{fig1-nebulae}, to extract two-dimensional spectra by integrating the associated voxels along the East direction (the spatial x-axis in Fig.~\ref{fig1-nebulae}). The x-axes in Fig.~\ref{fig1-spec} represent now the wavelength dimension and the y-axes correspond to the spatial y-axes in Fig.~\ref{fig1-nebulae} (with the same scale as in Fig.~\ref{fig1-nebulae}). Below each two-dimensional spectrum we also show the associated one-dimensional spectrum obtained by integrating all spatial pixels (black lines) and compare it to a scaled version of the original quasar spectrum (grey lines). 

Because we are using a two-dimensional projection of the 3D-mask, the number of integrated voxels, and therefore the noise level in the two-dimensional spectra, 
is constant across the wavelength direction at each fixed spatial position. 
The emergence of the extended line emission in this figure is clear, particularly in the integrated one-dimensional spectra. This figure also confirms that the detected emission for the smallest and more circular nebulae, e.g. \#11 and \#17, cannot be due to PSF removal artefacts, as the nebular spectrum (thick line) is always narrower than and strikingly different from the original quasar spectrum (thin lines).

\begin{figure}[t!]
\includegraphics[width=0.5\textwidth]{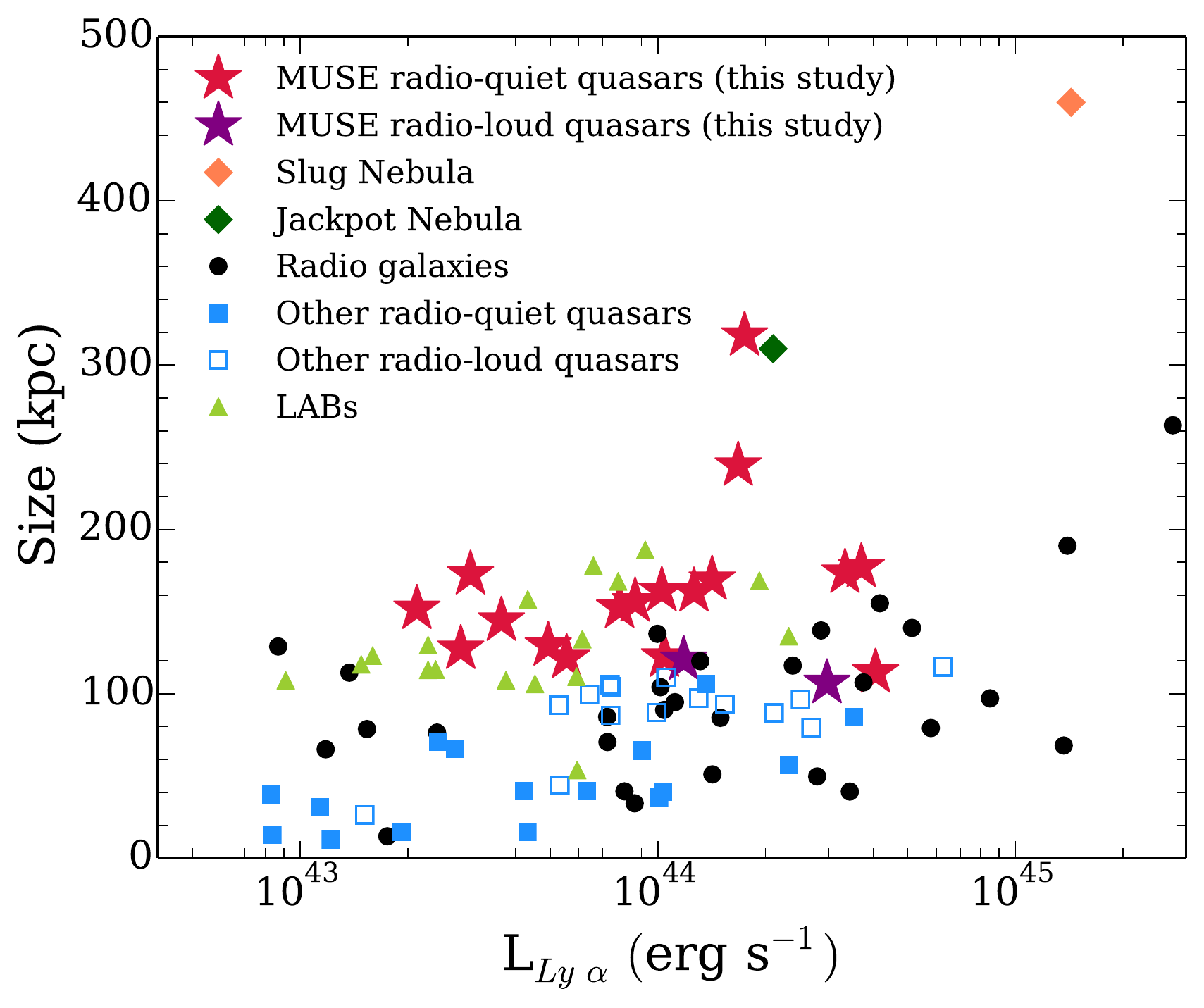}
 \caption{\lya  luminosities and maximum projected sizes of the giant \lya nebulae detected in this study around radio-quiet (red stars) and radio-loud (purple stars) quasars measured as the largest spatial projection of the CubExtractor "3D-mask" shown in Fig.~\ref{fig1-nebulae}. 
 For illustrative purposes, we overlay the results of other studies targeting different objects (see the legend in the image). Note however, that a direct comparison between the sizes and luminosities measured in MUSE datacubes and other
 studies is not possible because of the different sensitivity limits, redshifts, methods and definition of sizes. } 
 \label{fig3-LumSize}
\end{figure}

\begin{deluxetable*}{rlcccrccrr}{H}
\centering
\tablecolumns{8}
\tablewidth{0pc}
\tabletypesize{\footnotesize}
\tablecaption{Properties of the Detected Giant \lya Nebulae}
\tablehead{
           \colhead{Number}                                        & 
           \colhead{Object Name}                                   & 
           \colhead{$z_{\mbox{\tiny{QSO}}}$\,\tablenotemark{a}}    & 
           \colhead{$z_{\mathrm{Ly}\alpha}$\tablenotemark{b}}      & 
           \colhead{Size\,\tablenotemark{c}}                       & 
           \colhead{$\Delta\lambda$\,\tablenotemark{d}}            & 
           \colhead{Luminosity}                                    &
           \colhead{FWHM\,\tablenotemark{e}}                       &
           \colhead{C\,IV / \lya\,\tablenotemark{f}}               &                           
           \colhead{He\,II / \lya\,\tablenotemark{f}}              \\ 
           \colhead{}                                              & 
           \colhead{}                                              & 
           \colhead{}                                              & 
           \colhead{}                                              & 
           \colhead{(pkpc)}                                        & 
           \colhead{($\angstrom$)}                                 & 
           \colhead{(erg\,s$^{-1}$)}                               & 
           \colhead{(km\,s$^{-1}$)}                                &                           
           \colhead{($2\sigma$)}                                   &                           
           \colhead{($2\sigma$)}                                    
}   
\startdata
 1 & CTS G18.01   & 3.207  &  3.248  &  240 &  47.50  &  1.7$\times 10^{44}$  &  570  &  $<$0.04   &   $<$0.03   \\
 2 & Q0041-2638   & 3.036  &  3.078  &  170 &  23.75  &  2.9$\times 10^{43}$  &  320  &  $<$0.06   &   $<$0.02   \\
 3 & Q0042-2627   & 3.280  &  3.306  &  320 &  47.50  &  1.7$\times 10^{44}$  &  510  &  $<$0.01   &   $<$0.01   \\
 4 & Q0055-269    & 3.634  &  3.662  &  180 &  55.00  &  3.7$\times 10^{44}$  &  770  &     0.07   &   $<$0.03   \\
 5 & UM669        & 3.021  &  3.040  &  160 &  40.00  &  1.0$\times 10^{44}$  &  660  &  $<$0.05   &   $<$0.03   \\
 6 & J0124+0044   & 3.783  &  3.847  &  190 &  46.25  &  4.1$\times 10^{44}$  &  930  &     0.04   &   $<$0.03   \\
 7 & UM678        & 3.188  &  3.208  &  150 &  31.25  &  7.8$\times 10^{43}$  &  580  &  $<$0.06   &   $<$0.11   \\
 8 & CTS B27.07   & 3.132  &  3.155  &  160 &  35.00  &  1.0$\times 10^{44}$  &  590  &  $<$0.04   &   $<$0.06   \\
 9 & CTS A31.05   & 3.020  &  3.050  &  120 &  41.25  &  6.1$\times 10^{43}$  &  780  &  $<$0.09   &   $<$0.04   \\
10 & CT 656       & 3.125  &  3.159  &  130 &  31.25  &  2.8$\times 10^{43}$  &  640  &  $<$0.07   &   $<$0.06   \\
11 & AWL 11       & 3.079  &  3.118  &  130 &  30.00  &  4.9$\times 10^{43}$  &  670  &  $<$0.07   &   $<$0.04   \\
12 & HE0940-1050  & 3.050  &  3.091  &  170 &  45.00  &  1.4$\times 10^{44}$  &  660  &  $<$0.06   &   $<$0.01   \\
13 & BRI1108-07   & 3.907  &  3.935  &  160 &  53.75  &  1.2$\times 10^{44}$  &  760  &  $<$0.04   &   $<$0.04   \\
14 & CTS R07.04   & 3.351  &  3.368  &  170 &  40.00  &  3.3$\times 10^{44}$  &  660  &     0.05   &      0.01   \\
15 & Q1317-0507   & 3.701  &  3.720  &  140 &  33.75  &  3.6$\times 10^{43}$  &  560  &  $<$0.11   &   $<$0.09   \\
16 & Q1621-0042   & 3.689  &  3.704  &  120 &  31.25  &  5.5$\times 10^{43}$  &  550  &  $<$0.07   &   $<$0.05   \\
17 & CTS A11.09   & 3.121  &  3.150  &  150 &  28.75  &  2.1$\times 10^{43}$  &  490  &  $<$0.10   &   $<$0.05   \\
R1 & PKS1937-101  & 3.769 &  3.791  &  110  & 103.75  &  2.9$\times 10^{44}$  & 2120  &     0.11   &      0.12    \\
R2 & QB2000-330   & 3.759 &  3.788  &  120  &  52.50  &  1.2$\times 10^{44}$  & 1120  &     0.06   &   $<$0.06 
\enddata
\tablenotetext{a}{Systemic redshift of the quasar as in Table~\ref{tab1} derived from the peak of C\,IV line and corrected according to \citet{Schen2016_civ}.} 
\tablenotetext{b}{Measured from the flux-weighted centroid of the nebular \lya emission.}
\tablenotetext{c}{Maximum linear projected size measured from the spatial projection of the CubExtractor object segmentation mask ("3D-mask").} 
\tablenotetext{d}{Maximum spectral width of the "3D-mask" and width of the pseudo-NB used for the circularly-averaged SB profile.}
\tablenotetext{e}{Spatially averaged FWHM.}
\tablenotetext{f}{Limits on the line ratios correspond to 2$\sigma$.}
\label{tab2}
\end{deluxetable*}

\subsection{Morphology of the nebulae}

As is clear from Fig.~\ref{fig1-nebulae}, the nebulae have a large diversity of shapes and sizes. The smallest nebulae tend to be more symmetric and circular (with the exception of \#2) while the largest (\#1, \#3, extending up to 320\,pkpc) show evidence of filamentary structures and multiple components. The radio-loud quasar nebulae (R1 and R2) do not look particularly different from other nebulae in these images.
The brightest nebula (\#6) has a peculiar morphological structure with a sudden, steep decrease of the total flux around a distance of about 50\,pkpc from the quasar position. In some cases (e.g., \#7) the brightest part of the nebulae is clearly offset with respect to the quasar position and in general there seems to be a small offset between the nebula flux centroid and the quasar position. There is no evidence for "bipolar ionization cone illumination" patterns, although nebulae \#3 and \#13 shows hints of "bipolar" structure in their morphology.

We have defined the sizes of the nebulae as the maximum projected sizes of the 3D-masks (see Fig.~\ref{fig1-nebulae}) and have compared the sizes to the total luminosities as is commonly done in the recent literature, see Table~\ref{tab2} and 
Fig.~\ref{fig3-LumSize}. For illustrative purposes, we include in the same figure the results of other studies targeting radio-galaxies \citep{Reuland2003,Villar-Martin}, radio-loud quasars \citep{Heckman1991a}, Ly$\alpha$ blobs \citep{Matsuda2004,Prescott2009} and other radio-quiet quasars \citep{Bergeron1999, Christensen2006,North2012}. We stress that this comparison is only qualitative because these studies have different sensitivity limits, redshifts, methods and definitions of sizes.

\begin{figure*}[!ht]
\centering
\includegraphics[width=0.9\textwidth]{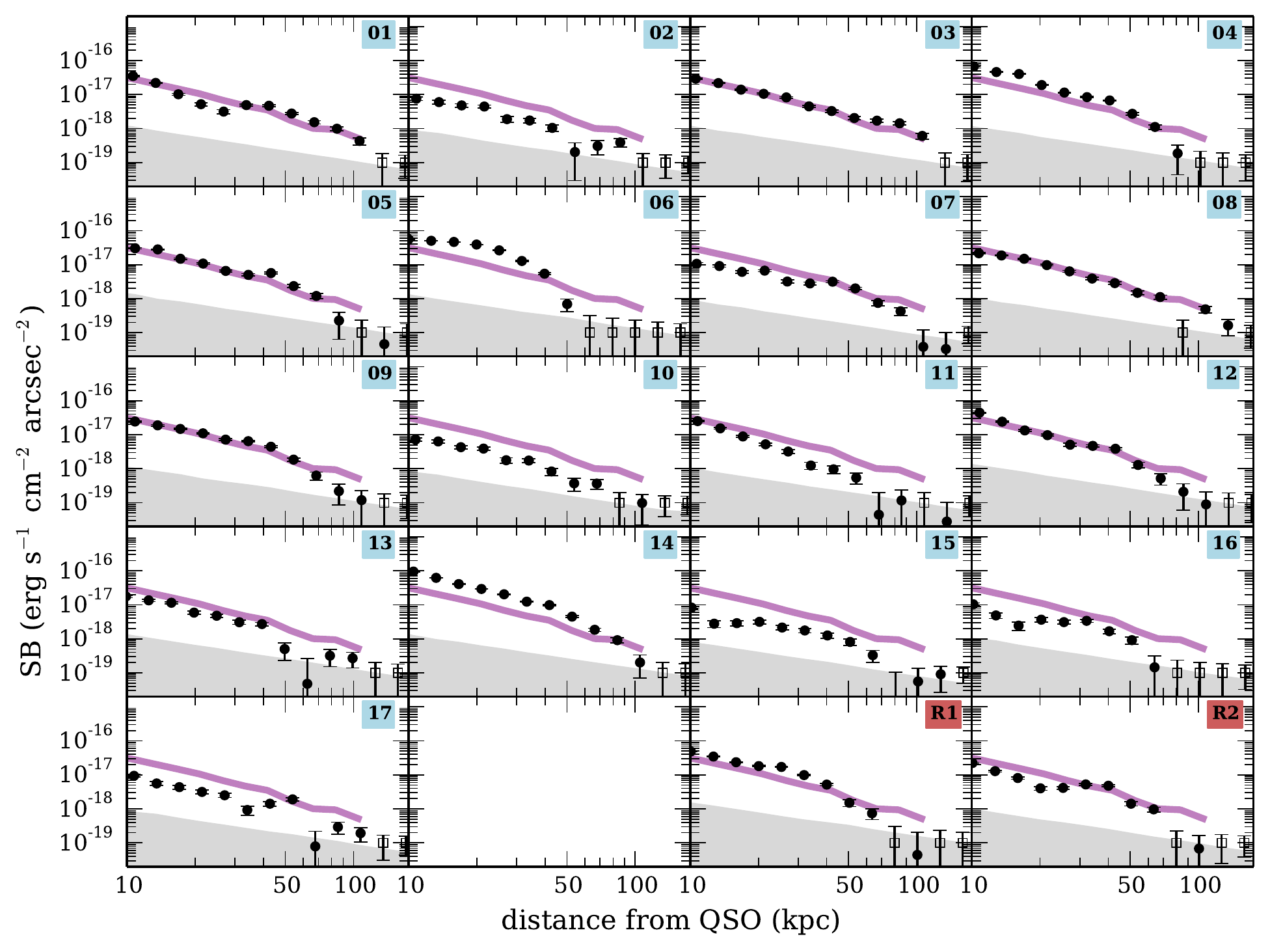}
\caption{Circularly averaged surface brightness (SB) profiles (black filled circles) as a function of projected distance from the quasar for each MUSE field (open squares indicate negative values). These SB profiles have been extracted from "fixed-width" pseudo-NB images instead of using the 3D-mask as in previous figures in order to avoid any possible SNR-threshold effect in the profile slopes (see text for discussion). Each point represents the SB level measured in a given annulus with bins uniformly spaced in a logarithmic scale and the error bars represent 1$\sigma$ errors. For reference, in each panel we show the average profile of all radio-quiet nebulae that is well fitted by a power law with a slope of $\approx -1.8$ (\textit{purple line}). The grey shaded area is a estimate of the 2$\sigma$ gaussian noise associated with the SB profiles. Interestingly, despite the very different morphologies, all nebulae show very similar circularly-averaged SB profiles (with the exception of \#6, see text for details).}
\label{SPplot_grid_individual}
\centering
\includegraphics[width=0.92\textwidth]{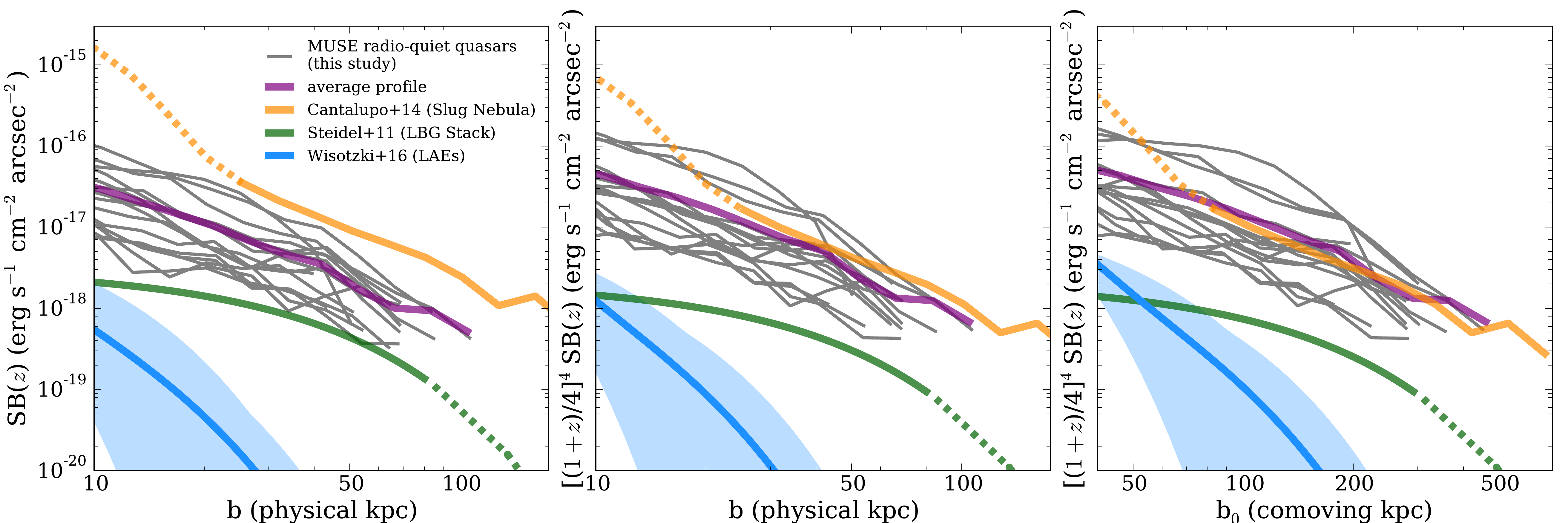}
 \caption{Comparison between the circularly averaged SB profiles of the MUSE detected \lya nebulae (grey lines are individual profiles and purple line is the average profile, as in Fig.~\ref{SPplot_grid_individual}) and other systems in the literature: Slug Nebula (orange line, the dashed line indicates the region contaminated by the quasar PSF), \lya haloes around individual galaxies detected in the MUSE-HDFS field (shaded blue area), and the LBG-stack of \citet{Steidel2011} (green line). The left-hand panel shows the observed SB as a function of the projected physical distance. In the central panel the x-axis is the same as before but the SB is redshift-dimming corrected. Finally, the right-hand panel shows the redshift-dimming corrected SB profiles as a function of comoving projected distance. Once corrected for the different redshifts, the slopes of the profiles of the MUSE detected nebulae are very similar to those of both the Slug Nebula and the LBG stack, suggesting a similar origin for all these different systems as discussed in detail in Section~\ref{discussion}. In particular, the redshift-corrected profile of the Slug Nebula is perfectly compatible with the typical profile of the MUSE-detected nebulae both in terms of slope and normalization.}
\label{SBplot-all}
\end{figure*}

\subsection{Surface brightness profiles}
\label{secSBprofs}

In this Section we investigate the average radial SB profile of the nebulae 
as a way of constraining the origin of the emission. Despite the different morphologies and clear asymmetries in some of the nebulae, we decided to use the standard approach in the literature of measuring circularly averaged SB profiles for both simplicity and ease of comparison with previous works. For similar reasons, we decided to use standard NB images derived from MUSE cubes, i.e.\ using a fixed width rather than the 3D-mask discussed above. In particular, we fix the width of these pseudo-NB images to the maximum spectral width of the nebulae as defined by the 3D-mask (see Table~\ref{tab2}). We note that the resulting profiles are noisier, especially at the edges of the nebulae, with respect to the images presented in Fig.\ref{fig1-nebulae}. However, the noise is better characterized and there are no effects due to SNR thresholding of the flux. We also computed and subtracted, if present, 
any residual background level from each pseudo NB\footnote{Such a residual background may be caused by faint background or foreground sources not removed or masked 
during continuum-subtraction, line emission or by residual cosmic rays.}. 
This residual background level was not higher than 1$\sigma$ of the flux pixel distribution in any of the case. Therefore, its contribution is small and only affects the profiles at larger distances from the quasars.  

The resulting circularly averaged SB profiles for each of the nebulae are presented as black lines in Fig.~\ref{SPplot_grid_individual}, while the purple line is the average profile that is well fitted by a power-law with slope of -1.8. The grey line is a 2$\sigma$ gaussian estimate of the noise associated with the profile. Surprisingly, despite the different morphologies, sizes and luminosities, all profiles look very similar to each other, including those of the radio-loud quasars, and are better represented by power-laws in the majority of the cases rather than exponential profiles (in Appendix~\ref{appB} we report the results of our fitting procedure including both power-law and exponential profiles). The most notable exception is \#6, which also has a peculiar morphology, as discussed in the previous Section. 

Fig.~\ref{SBplot-all} compares these profiles (black lines for individual profiles, purple line for the average profile, as in Fig.~\ref{SPplot_grid_individual}) with other studies. In particular, we show as a orange line the profile of the Slug Nebula \citep{C2014Natur}. We note that this profile was not PSF subtracted\footnote{PSF subtraction on Keck/LRIS images is particularly challenging given that the expected PSF shape may depend on the location on the detector.} and therefore we use as a dashed line style to indicate in the Slug Nebula the region of the profile that is contaminated by the quasar PSF. The blue line shows the average profile of \lya halos around \lya emitting galaxies in the deep MUSE observation of the HDFS \citep{Wisotzki2016} with the shaded blue area indicating the range of individual profiles. Finally, the green line is the stacked SB profile of NB images at \lya wavelengths of 92 Lyman Break Galaxies (LBGs) at $z\sim2.6$ obtained by \citet{Steidel2011}. 

Clearly, the profiles of the MUSE nebulae, presented in the leftmost panel, are very similar to the Slug Nebula despite the different luminosity and redshift and, somewhat more surprisingly, the profiles also have similar slopes as the LBG stacked profile found by \citet{Steidel2011}.On the other hand, the haloes of the average \lya emitters are clearly characterized by a different profile. As we will discuss in Section \ref{discussion}, these similarities may hint to a common fluorescent origin for quasar fluorescent nebulae and the LBG haloes in the fields observed by \citet{Steidel2011}. Once the different redshifts of the Slug Nebula and the nebulae found in this study are taken into account, we see in the middle and right panels that, despite its apparent larger size and luminosity, the Slug Nebula is perfectly compatible with the average detection obtained with MUSE.

\subsection{Kinematics}

Returning to the 3D-masks to define the location of the nebulae in the MUSE datacubes, we produced two-dimensional maps of the first and second moments of the flux distribution (in wavelength) to get an indication of the centroid velocity and width of the emission line for each spatial location. These resolved kinematic maps give us the opportunity to detect kinematic patterns, e.g. evidence for rotation, inflows or outflows and to identify kinematically distinct regions in the nebulae, if present (see e.g. \citealp{Prescott2015, Martin2015} for some examples). Because the shape of the \lya emission  from intergalactic gas cannot be simply modelled with a single analytic function, such as a Gaussian, we decided to follow this non-parametric approach of computing the moments of the flux distribution within the 3D-mask region. Restricting the analysis to the voxels associated with the CubExtractor 3D-masks significantly reduces the effect of the noise in this non-parametric approach.

\begin{figure*}
\centering
\includegraphics[width=0.9\textwidth]{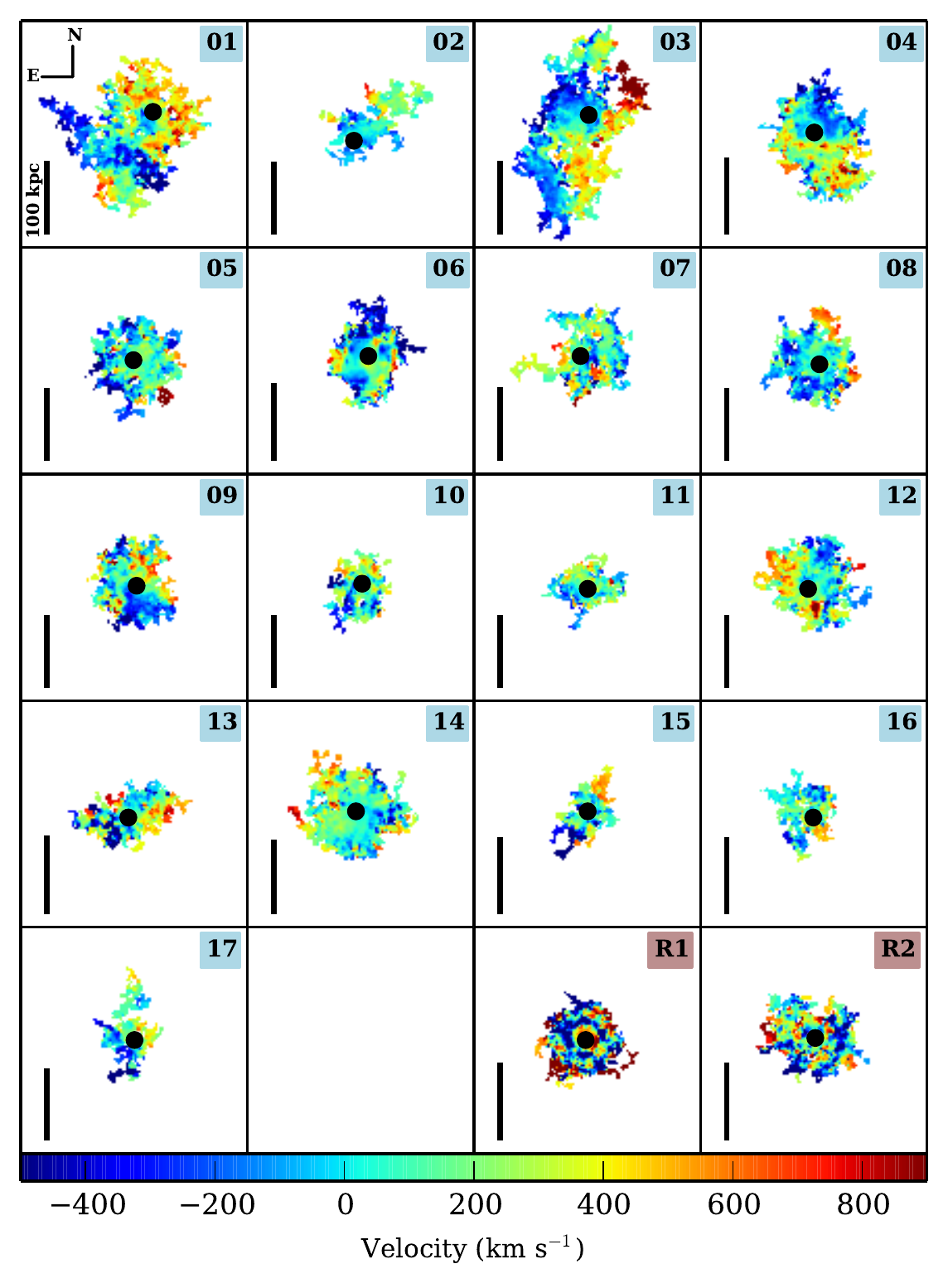}
 \caption{"Velocity maps" showing the flux-weighted velocity centroid shift with respect to the central wavelength of each nebula obtained from the first moment of the flux distribution (see text for details). The black circle indicates the position of the quasar. No clear "kinematic" patterns, e.g., rotation, inflow or outflows, are visible with the possible exception of \#15 (although we note that the \lya emission may be a poor tracer of the kinematics because of radiative transfer effects). Some nebulae show "kinematically" distinct filamentary structures (e.g. \#1 and \#3) that are remarkably coherent over very large spatial scales (about 100\,pkpc).}
 \label{fig4-Velocity}
\end{figure*}

In Fig.~\ref{fig4-Velocity}, we show the maps of the first moment of the flux distribution, i.e.\ the flux-weighted velocity centroid shift relative to the peak of the integrated \lya emission of each nebula. 
While some systems, e.g. \#15, show possible evidences of rotation in a disk-like structure, the majority of the nebulae do not show clear evidences from the \lya emission of rotation or other ordered kinematic patterns.
The largest nebulae show, instead, strikingly coherent kinematical structures over very large distances (e.g., \#1 and \#3).
We note that the \lya line is in general not the best indicator for kinematics because of radiative transfer effects, but these typically tend to disrupt coherency on large scales rather than enhance them \citep{C2005}. 

\begin{figure*}
\centering
\includegraphics[width=0.9\textwidth]{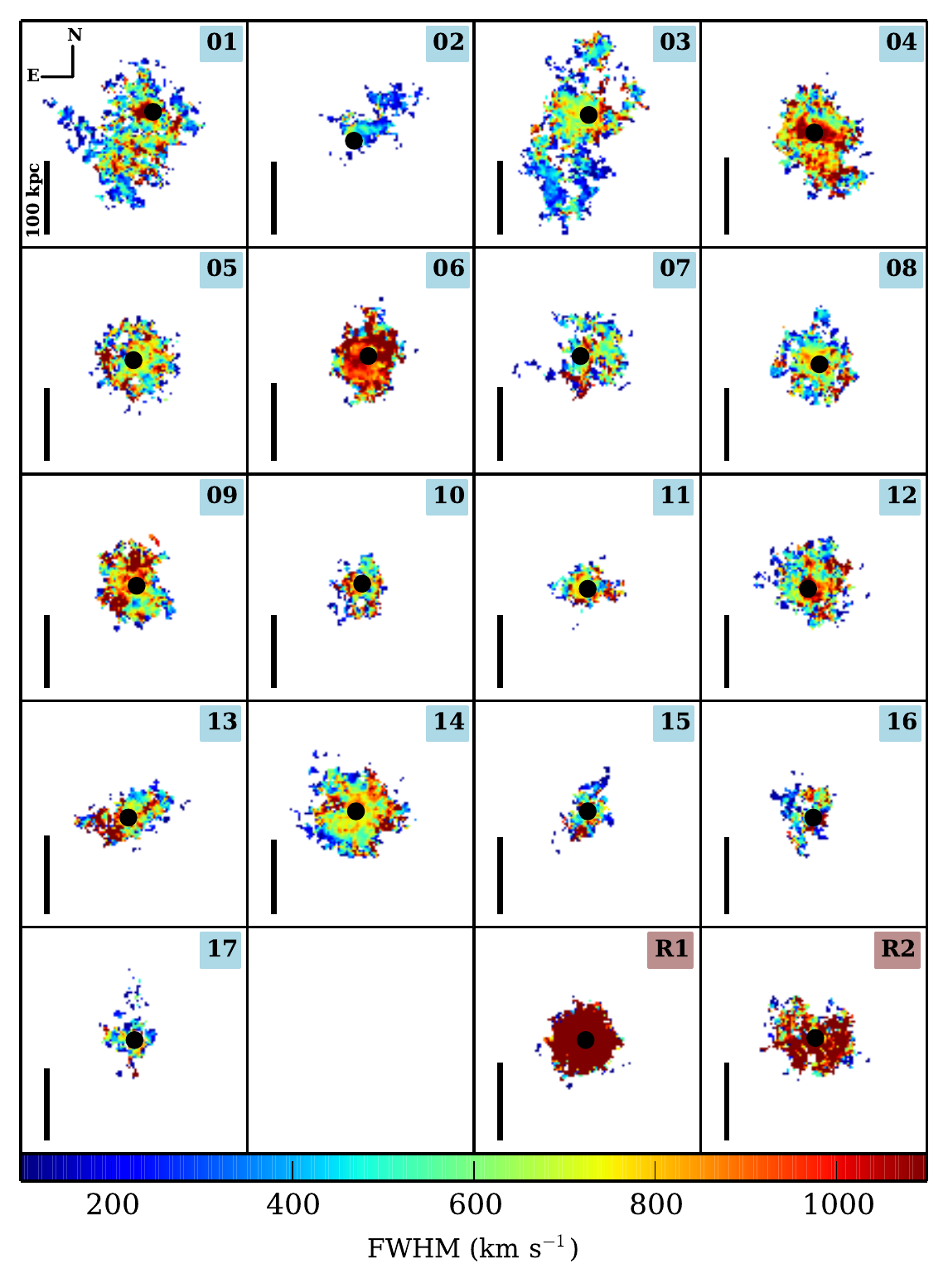}
 \caption{"Velocity dispersion maps" obtained from the second moment of the flux distribution for each of the MUSE nebulae. For consistency with previous works in the literature,  we show the Gaussian-equivalent FWHM (i.e., $2.35$ times the velocity dispersion). 
 These measurements are not corrected for the finite spectral resolution of the instrument (FWHM$\sim 170$ km s$^{-1}$) and therefore any value below this threshold should be considered as an upper limit.
 The black circle indicates the position of the quasar. We note that \lya emission may be broadened by radiative transfer effects, therefore the measurements presented here are not necessarily representative of the thermal or turbulent motion of the gas but they are typically an upper limit on these quantities. In relative terms, radio-quiet nebulae have significantly narrower lines with respect to radio-loud systems, in agreement with the expectations from the literature (the only exception is again nebula \#6).}
 \label{fig5-fwhm}
\end{figure*}

In Fig.~\ref{fig5-fwhm}, we present the maps of the second moment of the flux distribution, i.e.\ the velocity dispersion. For consistency with previous works in the literature 
we show the gaussian-equivalent FWHM derived by multiplying the second moment by $2.35$. Again, we stress that the \lya line may also be broadened by radiative transfer effects, so from this FWHM we cannot directly constrain the thermal properties of the gas. Nonetheless, the relative comparison between different objects is still informative. This figure clearly shows the main difference between radio-quiet and radio-loud systems: radio-quiet nebulae are narrower (500-700\,km\,s$^{-1}$) than radio-loud systems (FWHM\,$>$\,1000\,km\,s$^{-1}$) in agreement with previous results 
(e.g.,\citealp{Villar-Martin}).
There is only one radio-quiet nebula with broad line emission, i.e.\ \#6, and this is also the nebula that showed a clearly distinct SB profile (see Section~\ref{secSBprofs}). This is a striking confirmation that, even when ubiquitously detected, radio-quiet quasar nebula may have a different origin with respect to radio-loud systems. We will return to this point in our discussion below.

\subsection{Line ratios}
\label{lineratios}

The MUSE wavelength coverage allows us to search for other extended rest-frame UV lines, namely C\,IV/$\lambda1549\,\angstrom$ and  He\,II/$\lambda1640\angstrom$. These are typically the brightest UV lines after hydrogen \lya for AGN-photoionized nebulae and their line ratios have traditionally been used to constrain the origin and physical properties of the emitting gas \citep[see e.g.][]{Dey2005, Villar-Martin, Humphrey2008, Scarlata2009, Prescott2009, FabQSO2015, FabLAB2015}. In particular, the strengths of He\,II and C\,IV with respect to \lya may provide information on the ionisation parameter, gas density and metallicity, and eventually on the mechanism responsible for the gas emission. Giant \lya nebulae around high redshift radio galaxies typically show C\,IV and He\,II as well as other lines \citep{Villar-Martin}, which suggests a high metallicity for the gas and therefore an origin related to outflows rather than pristine intergalactic gas accretion. 

We again used the 3D-masks produced by CubExtractor to search for spatially coherent C\,IV and He\,II lines with the following strategy. Because the \lya emission peak may be shifted by radiative transfer effects with respect to other lines, we "scanned" around the expected location of these lines by shifting the \lya defined 3D-mask along the wavelength direction. We then compute the total flux of the voxels associated with the 3D-mask at each spectral location. Using the same three-dimensional aperture for \lya and the other lines guarantee that we do not have any aperture effects in the line ratios. However, because we expect the \lya emission to be always brighter (and more extended), this approach gives us very conservative limits in case of non-detections. In particular, we use the statistics of the "scan" to estimate the noise associated with the aperture photometry performed with the 3D-mask and we consider any line falling within $\pm3000$ km s$^{-1}$ from the expected line center with a flux higher than a $2\sigma$ detection. In case of a non detection, we use the $2\sigma$ as a limit. 

\begin{figure}
\centering
\includegraphics[width=0.5\textwidth]{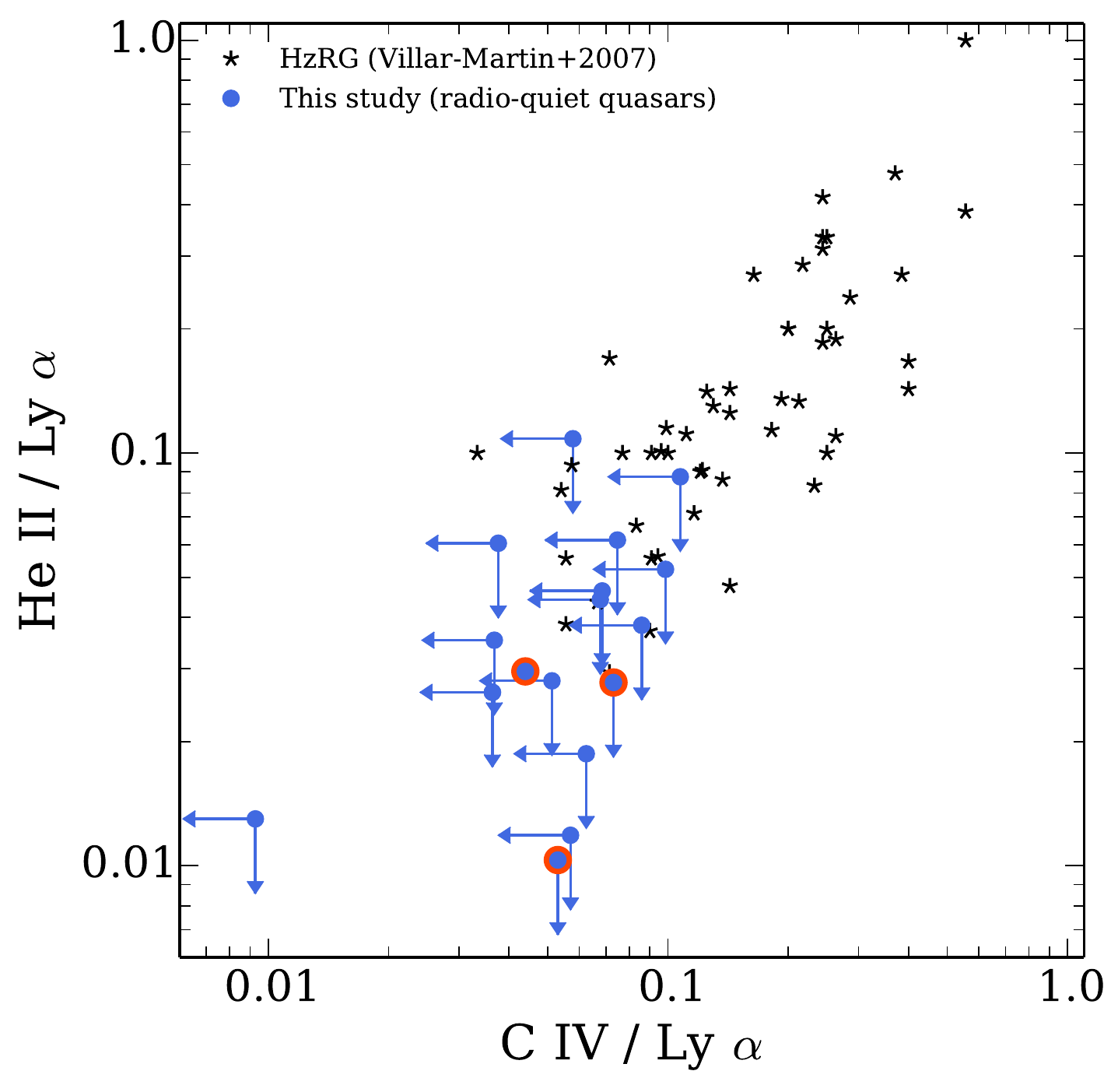}
 \caption{Comparison between the line ratios limit (2$\sigma$, blue arrows) and detections (red circles) for the MUSE nebulae (see Table~\ref{tab2}) and for radio-loud galaxies in the literature. The MUSE measurements have
 been obtained by matched three-dimensional aperture photometry as discussed in Section~\ref{lineratios}. The large majority of the nebulae are not detected (above 3$\sigma$) but the conservative
 limits estimated from our analysis place the MUSE nebulae in a different region of the line-ratio diagram with respect to radio-loud systems, suggesting a different origin for these sources, 
 e.g. in terms of metallicity.}
 \label{fig6-lines}
\end{figure}

The results are summarised in Table~\ref{tab2} and presented in Fig.~\ref{fig6-lines}, where we also compare them with previous studies of high-redshift radio galaxies (HzRG) from \citet{Villar-Martin}. In all but three radio-quiet systems we have detected neither C\,IV nor He\,II emission. The anomalous nebula \#6 (see above) is the only radio-quiet system with both (marginal) C\,IV and He\,II detections (at $2.2\sigma$ and $2.0\sigma$ respectively) while nebulae \#4 and \#14 only have detectable C\,IV (at $2.6\sigma$ and $2.8\sigma$ respectively). For all other nebulae we have only upper limits for C\,IV and  He\,II ($2\sigma$ in Fig.~\ref{fig6-lines}). Although we only have conservative limits in the majority of the cases, it is interesting to note that our nebulae seem to lie in a different part of the line ratio diagram with respect to radio-galaxy haloes, thus reinforcing the suggestion from the kinematic analysis of a different origin for these systems.
We leave a more detailed analysis of the detected sources in He\,II and/or in C\,IV and the implication of the line ratios for future work.

\section{Discussion}
\label{discussion}

\subsection{Why such a high detection rate compared to previous studies?}

In this study, we have shown that all brightest radio-quiet quasars in our sample, located at the redshift range $3<z<4$, are surrounded by giant ($>100\,$pkpc) \lya nebulae with a SB level above $\sim$10$^{-18}$~erg\,s$^{-1}$\,cm$^{-2}$\,arcsec$^{-2}$. This result seems to be in stark contrast with previous narrow-band and spectroscopic surveys for bright radio-quiet quasars at high-redshift. These studies have found detection rates of giant nebula less than 10\% using NB imaging at $2<z<3$ \citep{C2014Natur,Hennawi2015,Martin2014,Fab2016}, while spectroscopic surveys at any redshift have found exclusively smaller nebulae ($<60\,$pkpc) and only in about 50\% of the cases \citep{North2012,HennawiX2013,Fathivavsari+2015}. Because all these surveys - including our MUSE observations - reached a similar depth, the origin of this discrepancy must be searched for somewhere else. In particular, we explore in this Section the possibility that the different observational techniques may be playing an important role in producing such different results. As we will see, however, on the basis of current surveys we cannot completely exclude that a true evolution in the frequency of giant nebulae as a function of quasar redshift and luminosity may be present in the Universe based on current surveys, and future work will be needed to further explore this possibility. 

Let us assume that there is no redshift evolution in the intrinsic properties of giant \lya nebulae, around radio-quiet quasars and compare our results to NB surveys at $z\sim2-3$. As we have shown in Fig.~\ref{SBplot-all}, this assumption does not seem unreasonable: the Slug Nebula discovered in NB imaging \citep{C2014Natur} has a SB profile that is perfectly compatible with the MUSE detection once corrected for SB-redshift dimming (which should of course \emph{favour} detections at lower redshift). As discussed in the Introduction, there are two main effects that will reduce the detectability of such bright nebulae in NB imaging observations: i) Filter losses: if the nebular emission line falls at the edge or outside of the NB filter; and ii) PSF losses: if the quasar PSF hides the presence of the nebula or reduces our ability to detect extended emission. Spectroscopic surveys with a single slit orientation also suffer from slit losses, but they are not affected by filter losses.

From our MUSE \lya and quasar spectra we can estimate how many of our nebulae would have been missed by NB filter observations because of filter losses. For this thought experiment we will assume that quasars would have been selected for NB observations if their estimated systemic redshift from corrected broad-line emission (C\,IV in our case, but typically Mg\,II at $z\approx2$) would have placed the expected \lya emission at the peak of the NB filter transmission. In reality, the selection is made on a range of systemic redshifts (typically within 500\,km\,s$^{-1}$ from the filter center), but we will assume that this range is absorbed by the intrinsic scatter in the C\,IV-corrected redshift estimation ($\approx415$\,km\,s$^{-1}$ at 1$\sigma$, \citealp{Schen2016_civ}). Assuming a typical NB filter width of 3000\,km\,s$^{-1}$ (FWHM), we can notice from Fig.~\ref{fig1-spec} that two nebulae would have been totally missed by NB observations (\#2, \#6) while for the other five (\#1, \#10, \#11, \#12, \#16) the displacement between the \lya emission and the filter transmission peak would have resulted in a substantial flux loss that would likely have compromised the detectability of the nebulae, especially in the faintest but most extended regions. However, at least nine nebulae (\#3, \#4, \#5, \#7, \#8, \#9, \#13, \#14, \#15, \#17), which is at least 50\% of the sample, have \lya emission so close to the estimated systemic redshift that they would not have been missed by NB observations, including also the systemic redshift errors from C\,IV or Mg\,II. This factor is much larger than the 10\% detection rate estimated from NB imaging. Therefore, we conclude that filter losses cannot completely explain the discrepancy between MUSE detection rates and those of NB surveys, although they can certainly reduce the NB detection rate significantly. We note that radio-loud quasars typically have much broader \lya emission and therefore filter losses would be even less important for these systems. This may partially explain the higher detection rate of radio-loud nebulae versus radio-quiet ones. 

Regarding PSF losses, we notice that only about 40\% of the MUSE radio-quiet nebulae have a circular shape that could have been difficult to distinguish from the quasar PSF (\#5, \#8, \#12) or that are small and faint enough to make PSF-subtraction possibly challenging (\#10, \#15, \#16, \#17) for NB imaging or spectroscopy. At least 7 out of 17 nebulae (i.e.\ about 40\%) have clear asymmetric shapes or offsets from the quasar position that should have been easy to identify even with a very approximate, or absent, quasar PSF subtraction (\#1, \#2, \#3, \#6, \#7, \#13, \#14). The remaining 20\% of the sample (\#4, \#9, \#11) do not fall clearly in any of the previous categories. Despite the small number statistics, we do not think that PSF losses are the main
reason for the discrepancy with NB surveys, but they could contribute by reducing  the observed frequency of nebulae. This suggestion is reinforced by the fact that the two radio-loud nebulae discovered with MUSE (\#R1, \#R2) also have circular morphologies and that their detection rate should therefore not have been different from that for radio-quiet systems in NB surveys if PSF losses are important. Taken together, filter losses and PSF losses may result in a decrease by a factor 4, at most, in the detection rate of radio-quiet nebulae in NB observations, which still not enough to account for the discrepancy with MUSE data. 

The discrepancy with the detection rates measured in spectroscopic surveys can however be explained by a combination of slit losses and PSF losses. Given the complex and asymmetric morphology of a large fraction of MUSE nebulae, the probability that a single slit orientation would cover a 100\,pkpc patch of the extended emission is quite low. There are only a few cases in which the nebulae are symmetric enough to expect a detection over scales extending to 100\,pkpc (e.g., \#5), however the difficulty of estimating the exact PSF of the quasar through slit spectroscopy may result in a substantial reduction of the detection rate. This effect may also reduce the detectability of the inner parts of the nebulae to the level found by spectroscopic surveys \citep{North2012,HennawiX2013,Fathivavsari+2015}. Moreover, the availability of more spatial positions and the large FoV provided by MUSE with respect to a single slit will clearly facilitate the extraction and detection of extended emission around the quasars. 
Finally, it is easy to explain the discrepancy with other results based on IFS that have a much smaller FoV with respect to MUSE \citep[e.g.][]{Herenz2015,Christensen2006}: 
the minimum size of the MUSE detected nebulae is about 13 arcsec - which is much larger than the FoV of 8"$\times$8" of previous IFU surveys - and their SB profile is quite shallow in the central regions; 
because separate sky observations for sky-subtraction were not obtained in these surveys, such extended nebulae could have been removed as a sky feature (see e.g., Fig.7 in  \citealp{Herenz2015}).

If observational techniques and their limitations are not the only sources of the different detection rates, we should consider what other \emph{intrinsic} factors in the quasar sample selection may contribute to the difference. In particular, we notice that the MUSE quasar redshifts are systematically higher than those of quasars targeted by previous NB surveys but are compatible with some of the spectroscopic observations \citep[e.g.][]{North2012}. Similarly, the luminosities of the MUSE quasars are systematically higher than for  previous surveys because of the limited availability of quasars with redshifts falling in the same NB filters \citep[e.g.][]{C2014Natur, Fab2016} or quasar pairs for the largest spectroscopic survey to date \citep{HennawiX2013}. 
In particular, the quasars selected for this study are on average two absolute magnitude brighter (i-band rescaled to $z=2$) with respect to the sample of \citet{Fab2016}. 
We have explored if any of the main nebula parameters affecting the detection rate (e.g.\ size, luminosity, SB profile) depend on the quasar luminosity in our sample but found no significant correlations in all cases. However, we cannot exclude that a correlation may be present at much lower quasar luminosities with respect to the objects observed in this survey. Future studies extending the range of quasar luminosities probed with MUSE or extending to lower redshifts (e.g., with the Keck Cosmic Web Imager, \citealp{KeckCWI}) will help to resolve these questions.

\subsection{Origin of the nebulae and implications of our results}

Three mechanisms are able to produce extended and kinematically narrow \lya emission from cosmic hydrogen around a quasar, as observed in our study: 
i) recombination radiation due to quasar photo-ionisation \citep{C2005,Kollmeier2010,C2014Natur}; \
ii) "continuum-pumping" or \lya scattering from the quasar broad-line region \citep[see e.g.][]{C2014Natur};
iii) \lya collisional excitation (so called \lya "cooling radiation"; \citealp{Haiman+2000,Fardal2001,Dijkstra+2006,Cantalupo+2008},\citealp{RB2012} ).
All these mechanisms require the presence of "cool" gas around the quasars (i.e., with temperatures well below 10$^5$ K). In addition, recombination radiation and "continuum-pumping" also require that the gas is "illuminated" by the quasar or by some other bright source of UV photons. 

Estimating the possible contribution from \lya collisional excitation to the nebular emission is notoriously difficult, given the exponential dependence on temperature of the \lya collisional excitation rate \citep[see e.g.][]{Furlanetto2005,Cantalupo+2008}
and it requires radiation-hydrodinamical simulations with high spatial and temporal resolution in order to capture non-equilibrium ionisation effects in addition to accurate modelling of cooling and heating mechanism. 
On the other hand, if the gas is highly photo-ionised, \lya collisional excitation contribution is always negligible with respect to recombination radiation that has a weak dependence on temperature (in the range $10^3$ to $10^5$ K) and on ionisation state,
and it is therefore much easier to estimate. Given the presence of hyper-luminous quasars associated with our nebulae (by construction, in our survey), we think therefore that \lya collisional excitation is the least plausible of the three mechanism listed
above. If this was not the case, the small duty cycle of bright quasars at $z\sim3$ (e.g., 0.004$-$0.05, \citealp{Shen2007}) and our detection rate of 100\% would imply that giant and bright \lya nebulae should be about a hundred time more common than quasars 
at this redshift and this is certainly not observed\footnote{On the contrary, serendipitously discovered \lya nebulae are found later to be almost always associated with AGN \citep[e.g.][]{Prescott_LyaNeb}.}. 

A fluorescent origin - either from recombination radiation or "photon-pumping" - for the extended emission detected with MUSE implies that we are observing cosmic gas that is both relatively "cold" (T$\sim10^4$ K) and "illuminated" by a bright UV source, giving us the opportunity to constrain both the amount of cold gas around quasars and the geometry of their UV emission. In particular, from our 100\% detection rate of giant nebulae we can infer that "cold" gas is ubiquitous out to \emph{at least} 50 projected physical kpc from any hyper-luminous quasar at $z\approx3.5$, independent of the quasar emission opening angle. This distance covers about half of the virial radius for the expected dark matter halo associated with quasars at these redshifts (M$_{\mathrm{DM}}\sim10^{12.5}$M$_{\odot}$, independent of luminosity, see e.g.\ \citealp{Shen2007}), which instead are expected from numerical simulations to contain mostly hot, virialized gas at temperatures of T $>10^6$ K \citep[see][]{C2014Natur}. We stress that we do not see any kinematic signatures of extended disks or outflows, and therefore we believe that the origin of this gas is intergalactic. In some cases, we have detected cold gas up to a projected distance of about 200\,pkpc from the quasar (e.g. \#3), extending up to at least twice the virial radius of the typical dark matter haloes associated with quasars at this redshift.

Any constraints on the true total extent of the cold gas distribution are, however, degenerate with the geometry and opening angle of the quasar UV radiation. Turning this argument around, if the cold gas around quasars is extended (in all directions) up to the projected value that we measure in the images, then the quasar emission would have to be isotropic or covering at least 180 degrees if coming from a single "cone". Assuming instead that the dense gas that we can trace in emission always extends to at least 200\,pkpc from the quasars (using nebula \#3 as a reference, or assuming that we trace the same optically thick gas detected in absorption by \citealp{Prochaska2013}), then we will have a typical opening angle of about 25\% to 50\% of the total solid angle for biconical emission, or half of these values for a single "cone" emission. Future studies targeting both gas in emission and absorption around individual quasars (e.g., using quasar pairs) will help to break the degeneracy. 

What are the physical properties of this ubiquitous cold gas around bright quasars? From the SB profiles corrected for redshift-dimming and from the morphologies, we find that the observed emission properties of the nebulae are remarkably similar to that of the Slug Nebula \citep{C2014Natur}, despite the different redshifts, overall sizes and luminosities. On the one hand, our results show that these structures are common in the Universe around bright quasars and, on the other hand, all the implications of the Slug Nebula apply also in the case of the nebulae detected in this study.
In particular, the large values of the \lya SB and the lack of clear He\,II and C\,IV emission would imply large cold gas densities, unless a large clumping factor (with clump sizes of the order of few tens of pc, see \citealp{C2014Natur,FabQSO2015}) is invoked.

Remarkably, the LBG stacked SB profile obtained by \citet{Steidel2011} also has a very similar slope as to the Slug Nebula and MUSE detected nebulae. This profile is very different from the results of  other surveys targeting Ly$\alpha$ emitting galaxies both in large areas and using deep integrations (e.g., \citealp{Wisotzki2016}). How can we explain this apparent discrepancy? In view of our results, and considering that the quasar phase should only be a small part of the life of a galaxy, we must conclude that "cold" gas is always present around (massive) galaxies as well, and that the only difference for the appearance or not of a large nebula is the illumination factor provided by nearby or internal quasars. 
At this point it is worth noting that the LBGs observed by \citet{Steidel2011} are 
located in three fields of which one contains a hyper-luminous quasar (at the same redshift of the galaxies) and the other two contains two large overdensities of galaxies and AGN that might boost the ionizing UV background by a significant factor. This observation might then also be compatible with a fluorescent scenario, and it would imply the same origin for all giant extended emission around both quasars and galaxies.

\section{Summary and Conclusions}
\label{conclude}

In the last few years, direct Ly$\alpha$ imaging of fluorescent emission from intergalactic gas has begun to reveal giant cosmological structures around luminous quasars \citep[][e.g.]{C2014Natur,Martin2014,Hennawi2015} with high inferred gas density and relatively "cold" (T$\sim10^4$K) gas masses. Despite the high surface brightness of these sources, their observed detection rate from both NB imaging and spectroscopy surveys appeared very low, i.e.\ less than 10\%. If not affected by NB survey observational limitations, e.g. uncertainties in the quasar systemic redshift or quasar PSF-subtraction, such a low frequency may imply a small opening angle of the quasar emission, the lack of dense cold gas around the majority of quasars, or shorter quasar lifetime.

In this study, we have exploited the unique capabilities of the MUSE integral field instrument on the ESO/VLT to perform a blind survey for giant \lya nebulae around bright quasars at $z>3$ that does not suffer from most of the observational limitations of previous NB and spectroscopic surveys. In particular, we observed 17 of the brightest radio-quiet quasars (complemented by 2 radio-loud systems) at redshift $3<z<4$ for a total of 1h each reaching a SB limit of about 10$^{-18}$ erg s$^{-1}$ cm$^{-2}$ arcsec$^{-2}$ for a 1 arcsec$^{2}$ aperture (2$\sigma$) and unresolved line emission. After carefully correcting for residual systematics in the MUSE flat-fielding, removing the quasar PSF, and performing three-dimensional detection with CubExtractor and associated tools (Cantalupo, in prep.), we found that every observed quasar is associated at a high-significance level with giant \lya emission with projected linear sizes larger than 100\,pkpc.

Our detection rate of 100\% for giant \lya nebulae around radio-quiet quasars is in stark contrast with previous findings in the literature. While the asymmetric morphology of the MUSE nebulae may explain the discrepancy with spectroscopic surveys using a single slit position, the difference with the detection rate of NB surveys at $z\sim2$ cannot be completely explained by observational limitations alone. In particular, we estimated that the uncertainty in the quasar systemic redshifts, \lya emission line shifts and quasar-PSF subtraction can only account for a reduction of about a factor two in the detected frequency. Therefore, we cannot completely exclude the possibility that the frequency of giant nebulae changes with redshift and quasar luminosity.

The MUSE-detected nebulae present a large range in sizes and morphologies, ranging from circular nebulae with a projected diameter of about 100\,pkpc to filamentary structures with a projected linear size of 320\,pkpc. Despite these differences, the circularly averaged SB profiles show a strong similarity between all nebulae with very few exceptions and can be approximated by a power-law with a slope of about $\approx-1.8$. Remarkably, once corrected for redshift-dimming, the giant nebulae detected at $z\sim2$ using narrow-band imaging such as the Slug Nebula \citep{C2014Natur} are perfectly compatible with the average SB profile of the  nebulae reported here, both in terms of slope and normalization, suggesting a similar origin for these systems.

The lack of He\,II and C\,IV emission, the relatively narrow width of the \lya emission profiles, the intrinsic \lya SB of the nebulae, and the large extent of the emission strengthen the suggestion made by \citet{C2014Natur} and \citet{FabQSO2015} following the discovery of the Slug Nebula: a large fraction of the gas in and around the massive haloes hosting bright quasars could be in the form of kinematically quiet, dense ($n>1$ cm$^{-3}$), cold (T$\sim10^4$ K) small ($<20$ pc), and metal-poor ($Z<0.1\,Z_{\mathrm{sol}}$) clumps.  
Given these inferred clump sizes, current simulations of cosmological structure formation 
would unfortunately lack the spatial resolution (and, possibly, additional physics) to properly model this gas phase (see, e.g. \citealp{C2014Natur} for discussion\footnote{For similar discussions 
but more focused on absorption line studies around quasars, see \citealp{Prochaska2013,Fumagalli2014, Rahmati2015, Meiksin2015, FG2015}}).

In addition, our results imply that these gas clumps must be ubiquitous within at least 50\,pkpc from every bright quasar at $3<z<4$, independent of the quasar emission opening angle. This distance increases to at least 200\,pkpc if we assume that bright quasars have anisotropic emission and that variation in emission geometry and opening angles are the origin for the different sizes of the MUSE detected nebulae. 

This first exploratory MUSE survey for quasar fluorescent emission demonstrates that even relatively short integration times (1h) with MUSE
coupled with advanced data reduction and analysis tools are able to provide a three-dimensional view of the intergalactic gas \emph{in emission}  
around \emph{any} bright quasar at high redshift. Our study paves the way for future surveys that will both increase the statistical sample in terms of the explored parameter space
(e.g., quasar redshift, luminosity, environment or other emission properties) and perform a deeper and more detailed analysis on individual systems
(e.g., deriving the spatially resolved density, temperature, kinematics and "clumpiness" of the gas). Combined with a new generation of theoretical and numerical models, 
this new observational probe will provide a new window on both cosmic structure formation and the emission properties of the brightest quasars
in the high-redshift Universe.

\section*{Acknowledgement}

This work has been supported by the Swiss National Science Foundation. JS acknowledges partial support by the European Union's Seventh Framework Programme (FP7/2007-2013) / ERC Grant agreements 278594-GasAroundGalaxies. JR acknowledges support from the ERC starting grand 336736-CALENDS.

\appendix
\section{A. Comparison between optimally-extracted and fixed-width, pseudo-Narrow-Band images}
\label{appA}

\begin{figure}[h!]
\centering
\includegraphics[width=0.8\textwidth]{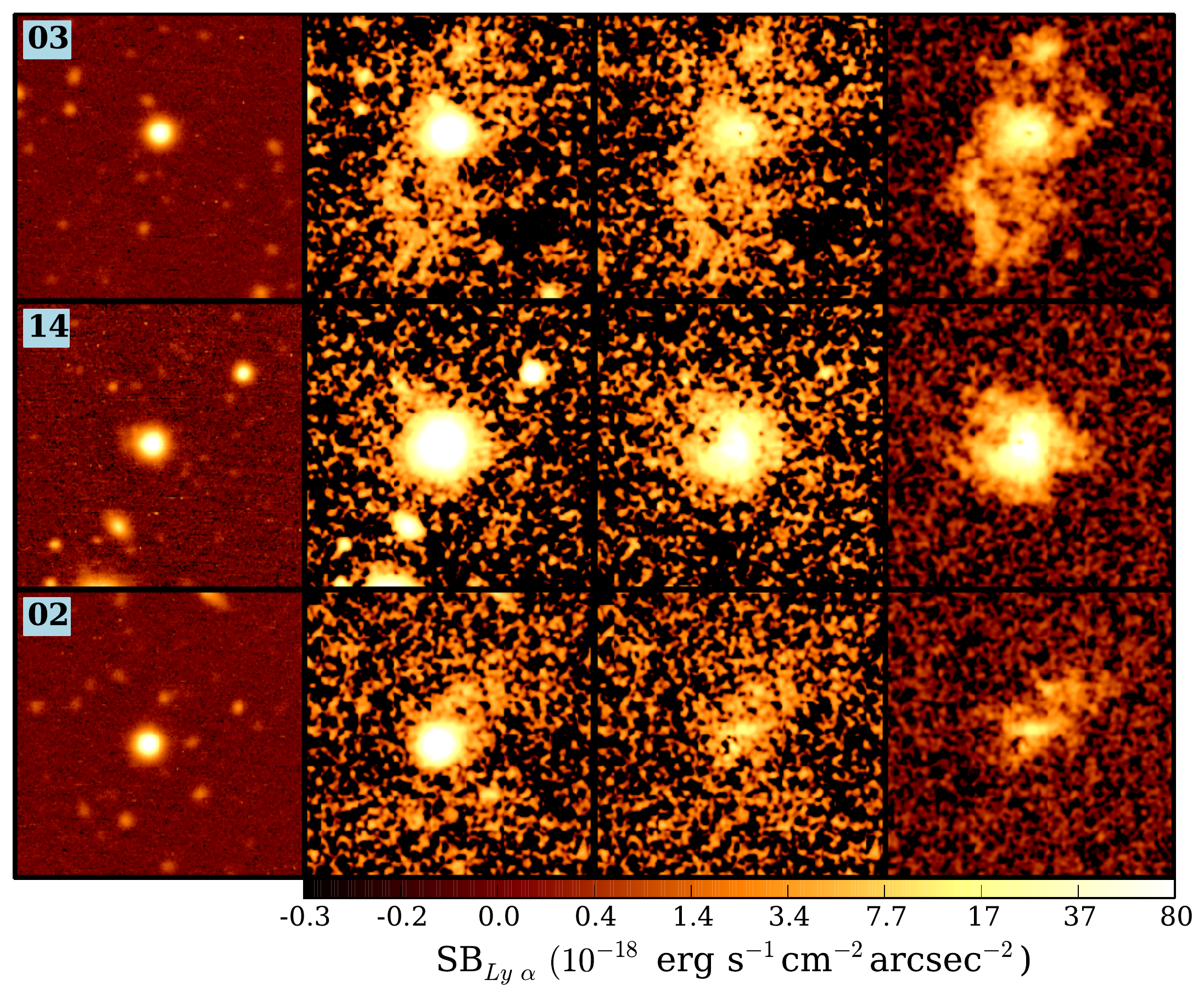}
 \caption{From the left to the right for each row: i) white-light images, ii) pseudo-NB, iii) pseudo-NB including continuum and quasar PSF subtraction, iv) optimally 
extracted images. The labels on the left-hand panels refer to the nebula number as in the main text. The spectral width of the pseudo-NB image for each of these nebulae is reported in Table~\ref{tab2}.}
 \label{fig-app}
\end{figure}

In this Section we show for illustrative purposes the comparison between the optimally-extracted (see Fig.~\ref{fig1-nebulae}) and fixed-width NB images (with and without PSF subtraction) obtained from MUSE datacubes using CubExtractor (Cantalupo, in prep.)
for three representative nebulae with different sizes and morphologies. In Fig.\ref{fig-app}, we show from the left to the right the following images: i) broad-band image (with the width of $\approx2600\,\angstrom$), ii) pseudo-NB, iii) pseudo-NB without continuum objects and quasar PSF, iv) optimally 
extracted. The labels on the left-hand panels refer to the nebula number as in the main text. The spectral width of the pseudo-NB image for each of these nebulae is reported in Table~\ref{tab2}. On one hand these images demonstrate the advantage 
of three-dimensional extraction and detection on MUSE datacubes. On the other hand, they show that the largest and the more asymmetric nebulae could have been easily detected also by traditional NB imaging
if a perfectly matched NB filter to the nebular spectral width would have been available.

\section{B. Surface brightness profile fitting}
\label{appB}

In this Section, we report the result of our fitting procedure of the circularly averaged SB profile for each nebula presented in the main text.
In particular, we have used a power-law and an exponential analytical profile described, respectively, as:

\begin{equation}
 \mbox{SB}(r) = C_{p}\,r^{\alpha}
\end{equation}

and

\begin{equation}
 \mbox{SB}(r) = C_{e}\,\exp^{-r/r_h}
\end{equation}

\begin{deluxetable}{rlrcrcrr}[h!]
\centering
\centering
\tablecolumns{8}
\tablewidth{0pc}
\tabletypesize{\footnotesize}
\tablecaption{Results of the SB fitting}
\tablehead{
           \colhead{Number}                                & 
           \colhead{Object Name}                           & 
           \colhead{$\alpha$\,\tablenotemark{a}}           & 
           \colhead{$\log10( C_{p,r10}$)\,\tablenotemark{b}} & 
           \colhead{$r_{h}$\,\tablenotemark{c}}            & 
           \colhead{$\log10 (C_{e,r10}$)\,\tablenotemark{d}} &
           \colhead{$(\chi^2_{p})$\,\tablenotemark{e}}         &
           \colhead{$(\chi^2_{e})$\,\tablenotemark{f}}         \\ 
           \colhead{}                                      & 
           \colhead{}                                      & 
           \colhead{}                                      & 
           \colhead{}                                      & 
           \colhead{(pkpc)}                                & 
           \colhead{}                                      & 
           \colhead{}                                      &                           
           \colhead{}                                                  
}   
\startdata
 1 & CTS G18.01   & -1.62  &  -16.54  &  60.55  &  -16.84  &   8.6  &  42.2  \\
 2 & Q0041-2638   & -1.62  &  -16.97  &  53.30  &  -17.23  &   1.2  &   1.7  \\
 3 & Q0042-2627   & -1.61  &  -16.47  &  60.84  &  -16.77  &   0.6  &  15.2  \\
 4 & Q0055-269    & -2.12  &  -16.06  &  31.78  &  -16.29  &   7.7  &  10.0  \\
 5 & UM669        & -1.67  &  -16.40  &  43.48  &  -16.61  &   3.1  &   8.0  \\
 6 & J0124+0044   & -2.83  &  -15.72  &  20.69  &  -15.99  &  46.5  &  16.1  \\
 7 & UM678        & -1.43  &  -16.83  &  55.66  &  -17.03  &   2.7  &   1.5  \\
 8 & CTS B27.07   & -1.74  &  -16.50  &  56.10  &  -16.82  &   2.3  &   8.5  \\
 9 & CTS A31.05   & -1.78  &  -16.42  &  38.88  &  -16.62  &   5.2  &   1.0  \\
10 & CT 656       & -1.78  &  -16.97  &  40.12  &  -17.18  &   1.3  &   0.7  \\
11 & AWL 11       & -2.43  &  -16.50  &  26.02  &  -16.75  &   0.2  &   5.7  \\
12 & HE0940-1050  & -2.16  &  -16.29  &  33.28  &  -16.56  &   2.5  &  16.7  \\
13 & BRI1108-07   & -1.87  &  -16.61  &  53.74  &  -16.97  &   1.1  &   3.4  \\
14 & CTS R07.04   & -2.14  &  -15.88  &  37.39  &  -16.18  &  10.5  &  18.1  \\
15 & Q1317-0507   & -1.47  &  -17.14  &  87.87  &  -17.48  &   1.9  &   8.1  \\
16 & Q1621-0042   & -1.11  &  -17.12  &  52.44  &  -17.21  &   3.2  &   4.6  \\
17 & CTS A11.09   & -1.54  &  -16.99  &  63.52  &  -17.27  &   2.2  &   4.9  \\                
R1 & PKS1937-101  & -2.16  &  -16.15  &  29.71  &  -16.35  &  18.3  &   2.9  \\
R2 & QB2000-330   & -1.44  &  -16.74  &  46.73  &  -16.89  &   5.6  &  11.5 
\enddata

\tablenotetext{a}{The slope from the power law fit to the surface brightness profile of each nebula.} 
\tablenotetext{b}{Normalization parameter for the power law fit at the radius of $r=10\,$pkpc.}
\tablenotetext{c}{The scale length from the exponential law fit to the surface brightness profile of each nebula.} 
\tablenotetext{d}{Normalization parameter for the exponential law fit at the radius of $r=10\,$pkpc.}
\tablenotetext{e}{$\chi^{2}$ statistics for the power law fit.}
\tablenotetext{f}{$\chi^{2}$ statistics for the exponential law fit.}

\label{tab3}
\end{deluxetable}

Table~\ref{tab3} summarize the results of the fitting procedure, including $\chi^2$ statistics. For each profile we have only used the data points with a SNR above $2$. 

\vspace{1cm}

\bibliographystyle{apj}
\bibliography{apj-jour,refs}

\end{document}